%% file: main.tex
\documentclass{article}

\usepackage{times}              % Font family selection
\usepackage{latexsym}
\usepackage{amsmath}
\usepackage{amsfonts}
\usepackage{amsthm}
\usepackage{epsfig}
\usepackage{cite}
\usepackage{graphicx}
\usepackage{fancyheadings}
\usepackage{algorithmic}
\usepackage{fullpage}

\newcommand{\prefix}{\textnormal{\sc pre}}
\newcommand{\suffix}{\textnormal{\sc suf}}

\newcommand{\sto}{\textnormal{s.t.}}

\newcommand{\opt}{\textnormal{\sc opt}}

\newcommand{\p}{\mathbf{P}}

\newcommand{\be}{\begin{enumerate}}
\newcommand{\ee}{\end{enumerate}}
\newcommand{\bi}{\begin{itemize}}
\newcommand{\ei}{\end{itemize}}
\newcommand{\beq}{\begin{equation}}
\newcommand{\eeq}{\end{equation}}

\newcommand{\bp}{\begin{proof}}
\newcommand{\ep}{\end{proof}}
\newcommand{\bcor}{\begin{cor}}
\newcommand{\ecor}{\end{cor}}
\newcommand{\bthm}{\begin{thm}}
\newcommand{\ethm}{\end{thm}}
\newcommand{\blmm}{\begin{lmm}}
\newcommand{\elmm}{\end{lmm}}
\newcommand{\bdefn}{\begin{defn}}
\newcommand{\edefn}{\end{defn}}
\newcommand{\bprop}{\begin{prop}}
\newcommand{\eprop}{\end{prop}}
\newcommand{\bconj}{\begin{conj}}
\newcommand{\econj}{\end{conj}}
\newcommand{\bopm}{\begin{opm}}
\newcommand{\eopm}{\end{opm}}
\newcommand{\brmk}{\begin{rmk}}
\newcommand{\ermk}{\end{rmk}}

\newcommand{\suchthat}{\ | \ }

\newcommand{\mv}[1]{\mathbf{#1}}

\theoremstyle{plain}                   % default
\newtheorem{thm}{Theorem}[section]
\newtheorem{lmm}[thm]{Lemma}
\newtheorem{prop}[thm]{Proposition}
\newtheorem{cor}[thm]{Corollary}

\theoremstyle{definition}              % Examples and all

\newtheorem{opm}{Open Problem}
\newtheorem{conj}[thm]{Conjecture}
\newtheorem{ex}[thm]{Example}

\newtheorem{defn}[thm]{Definition}

\newtheorem{rmk}[thm]{Remark}

\newcommand{\bbox}{
\begin{center}
\begin{tabular}{|c|}
\hline
}
\newcommand{\ebox}{
\\
\hline
\end{tabular}
\end{center}
}

\newlength{\toppush}
\begin{document}

%%%%%%%%%%%%%%%%%%%%%%%%%%%%%%%%%%%%%%%%%%%%%%%%%%%%%%%%%%%%%%%%%%%%%%%%%%%%%%%
%% the title
\title{Analyzing Nonblocking Switching Networks using Linear 
       Programming (Duality)}

\author{
Hung Q. Ngo, Atri Rudra\thanks{Atri Rudra was supported in part by NSF CAREER
Award CCF-0844796.}, Anh N. Le, Thanh-Nhan Nguyen\\
Computer Science and Engineering\\
The State University of New York at Buffalo\\
Email: \texttt{\{hungngo, atri, nguyen9, anhle\}@buffalo.edu}}

\maketitle

%%%%%%%%%%%%%%%%%%%%%%%%%%%%%%%%%%%%%%%%%%%%%%%%%%%%%%%%%%%%%%%%%%%%%%%%%%%%%%%
%% The Body

%\input{abstract}

\begin{abstract}
The main task in analyzing a switching network design (including circuit-, 
multirate-, and photonic-switching) is to determine the minimum 
number of some switching components so that the design is non-blocking in 
some sense (e.g., strict- or wide-sense). We show that, in many cases, this 
task can be accomplished with a simple two-step strategy: 
(1) formulate a linear program whose optimum value is a bound for the 
minimum number we are seeking, and 
(2) specify a solution to the dual program, whose objective value by weak 
duality immediately yields a sufficient condition for the design to be 
non-blocking.

We illustrate this technique through a variety of examples, ranging from 
circuit to multirate to photonic switching, from unicast to $f$-cast and 
multicast, and from strict- to wide-sense non-blocking.  The switching 
architectures in the examples are of Clos-type and Banyan-type, which are the 
two most popular architectural choices for designing non-blocking switching 
networks. 

To prove the result in the multirate Clos network case, we formulate a new 
problem called {\sc dynamic weighted edge coloring} which generalizes the 
{\sc dynamic bin packing} problem.  We then design an algorithm with 
competitive ratio $5.6355$ for the problem. The algorithm is analyzed using 
the linear programming technique. 
A new upper-bound for multirate wide-sense non-blocking Clos networks follow, 
improving upon a decade-old bound on the same problem. 
\end{abstract}


%%%%%%%%%%%%%%%%%%%%%%%%%%%%%%%%%%%%%%%%%%%%%%%%%%%%%%%%%%%%%%%%%%%%%%%%%%%%%%%
\section{Introduction}
\label{section:intro}
%%%%%%%%%%%%%%%%%%%%%%%%%%%%%%%%%%%%%%%%%%%%%%%%%%%%%%%%%%%%%%%%%%%%%%%%%%%%%%%

The two most important architectures for designing non-blocking
switching networks are Clos-type \cite{clos:network} and Banyan-type 
\cite{goke-1973}.
The Clos network not only played a central role in classical
circuit-switching theory \cite{MR35:1408,MR2000g:94065}, 
but also was the bedrock of multirate switching 
%\cite{Verm236, melenturner2003,
%MR92c:68010, DBLP:journals/ton/LiewNC98, 
%ngo-vu-2002, ngo-2002-1, MR98e:68008} 
\cite{melenturner2003,
MR92c:68010, DBLP:journals/ton/LiewNC98, 
ngo-vu-2002, ngo-2002-1, MR98e:68008} 
(e.g., in time-divisioned switching environments where connections
are of varying bandwidth requirements),
and photonic-switching
%\cite{ngo-pan-qiao-2006:ton,yang3,yang-2000,wilfong99wdm, haxell2002, 
%MR1755519, rasala-stoc2000}.
\cite{ngo-pan-qiao-2006:ton, haxell2002, MR1755519, rasala-stoc2000}.
The Banyan network is isomorphic to various other ``bit-permutation'' networks
such as Omega, baseline, etc.,
\cite{MR1700855}; 
they are called {\em Banyan-type} networks
and have been used extensively in designing electronic and optical switches,
as well as parallel processor architectures \cite{IB-D973021}.
In particular, the multilog design which involves the vertical stacking of 
a number of inverse Banyan planes has been used 
in circuit- and photonic-switching environments
because they have small depth ($\log N$), 
self-routing capability, and absolute signal loss uniformity
%\cite{Lea90,vaez-lea-2000,Hwang98,Shy912,
%maier-pattavina-2001,Shy91,966007,jiang-2009}.
\cite{Lea90,vaez-lea-2000,Shy912, maier-pattavina-2001,Shy91}.

In analyzing Clos networks, the most basic task is to
determine the minimum number of middle-stage crossbars
so that the network satisfies a given nonblocking condition.
This holds true in space-, multirate-, and photonic-switching,
in unicast, $f$-cast and multicast, and broadcast traffic patterns,
and in all nonblocking types (strict-sense, wide-sense, and rearrangeable).
Similarly, analyzing multilog networks often involves determining
the minimum number of Banyan planes so that the network satisfies
some requirements.
This paper shows that a simple and effective linear programming (LP) based
two-step strategy can be employed in the analysis:
\bi
\item First, the minimum value we are seeking 
(e.g., the number of middle-stage
crossbars in a Clos network or the minimum number of 
Banyan planes in a multilog network) is
upper-bounded by the optimum value of a linear program (LP)
of the form $\max\{\mv c^T\mv x \suchthat \mv{Ax\leq b, x\geq 0}\}$.
The maximization objective is often required by worst-case analysis,
such as the maximum number of middle-stage crossbars in a Clos network 
which is {\em in}sufficient to carry a new request.
The constraints of the LP are used to express the fact that no 
input or output can
generate or receive connection requests totaling more than its capacity.
\item Second, by specifying {\em any} feasible solution, say $\mv y^*$,
to the dual program $\min\{\mv b^T\mv y \suchthat \mv A^T\mv y \geq \mv c\}$,
and applying weak duality we can use the dual-objective value 
$\mv b^T \mv y^*$ as an upper bound for the minimum value being sought. 
\ei

In some cases, we may not need the second step
because the primal LP is small with only a few variables.
In most cases, however, the LP and its dual are very general, 
dependent on various parameters of the switch design.
% (number of inputs and switching elements' dimensions, for example).
In such cases, it would be difficult to come up with a 
primal-optimal solution.
Fortunately, we can supply a dual-feasible solution to quickly 
``certify'' the bound.
%In fact, as we shall see later in this paper, we might need to
%specify a family of dual-feasible solutions and choose the best one
%(i.e. the one giving the smallest objective value) 
%in accordance with the parameters of the problem at hand.

The LP-duality technique was first used in our recent paper
\cite{ngo-cocoon2008} to analyze the (unicast) strictly nonblocking 
multilog architecture in the photonic-switching case, subject to
general crosstalk constraints.
This paper demonstrates that the technique can be applied
to a wider range of switching network analysis problems.
%We use the Clos and multilog networks as 
%canonical examples to illustrate our analysis technique. 
Our main contributions are as follows.
First, we formulate a new problem called {\sc dynamic weighted edge coloring} 
({\sc dwec}) of graphs, which generalizes the classic 
{\sc dynamic bin packing} problem \cite{MR697157}
and the routing problem for multirate widesense nonblocking Clos networks.
Using the LP-technique, we design an algorithm with competitive ratio
$5.6355$. 
%We also show that no algorithm can have competitive ratio
%better than $4-O(\log n/n)$.
%New lower and upper bounds for the multirate Clos network problem follow.
A new upper bound for the multirate Clos network problem follows.
Since {\sc bin packing} and its variations have been very useful
in both theory and practice, 
we believe that {\sc dwec} and our results on it are of independent interest.
Second, we use the LP-technique to prove general sufficient conditions for
the multilog network to be $f$-cast nonblocking under the
so-called {\em window algorithm}, both under the link-blocking model
and the crosstalk-free model.
To the best of our knowledge, these are the first $f$-cast results for
the multilog design.
We show that many known results are immediate corollaries of these
general conditions.

The rest of this paper is organized as follows.
Section \ref{sec:prelim} presents notations and terminologies.
Section \ref{sec:clos} illustrates the strength of the LP-duality
technique on analyzing several problems on the Clos networks.
The {\sc dwec} problem is also defined and analyzed.
Section \ref{sec:fcast-WSNB} proves non-blocking
results for $f$-cast multilog architecture.
Section \ref{sec:cf-fcast-WSNB} does the same with the crosstalk constraint.
%Section \ref{sec:remarks} concludes the paper with a few remarks.
%\input{prelim}

%%%%%%%%%%%%%%%%%%%%%%%%%%%%%%%%%%%%%%%%%%%%%%%%%%%%%%%%%%%%%%%%%%%%%%%%%%%%%%%
\section{Preliminaries}
\label{sec:prelim}
%%%%%%%%%%%%%%%%%%%%%%%%%%%%%%%%%%%%%%%%%%%%%%%%%%%%%%%%%%%%%%%%%%%%%%%%%%%%%%%

Throughout this paper, for any positive integers $k,d$, let
$[k]$ denote the set $\{1,\dots,k\}$,
%$[k] = \{1,\dots,k\}$,
$\mathbb{Z}_d$ denote the set $\{0,\dots,d-1\}$ which can be thought
         of as $d$-ary ``symbols,"
%$\mathbb{Z}_d = \{0,\dots,d-1\}$,
let $\mathbb{Z}_d^k$ denote the set of all $d$-ary strings of length $k$,
$|\mv s|$ the length of any $d$-ary string $\mv s$ (e.g., $|3142| = 4$),
and $\mv s_{i..j}$ the substring $s_i\cdots s_j$
of a string $\mv s=s_1\dots s_l \in \mathbb{Z}_d^l$
(if $i>j$ then $\mv s_{i..j}$ is the empty string).

%%%%%%%%%%%%%%%%%%%%%%%%%%%%%%%%%%%%%%%%%%%%%%%%%%%%%%%%%%%%%%%%%%%%%%%%%%%%%%%
\subsection{Switching environments}

Consider an $N \times N$ switching network, i.e. a switching network
with $N$ inputs and $N$ outputs.
There are three levels of nonblockingness of a switching network.
A network is {\em rearrangeably nonblocking} (RNB) if
it can realize any one-to-one
mapping between inputs and outputs simultaneously;
it is {\em widesense nonblocking} (WSNB) if a new request from a free input
to a free output can be realized without disturbing existing connections,
as long as all requests are routed according to some algorithm;
finally, it is {\em strictly nonblocking} (SNB) if a new request
from a free intput to a free output can always be routed no matter how
existing connections were arranged.
In the multicast case, RNB, WSNB, and SNB are defined similarly.
%There are three nonblockingness-degrees of a switching network:
%rearrangeably nonblocking (RNB), widesense nonblocking (WSNB),
%and strictly nonblocking (SNB).
The reader is referred to \cite{MR2000g:94065} for more details on 
non-blocking concepts.

In circuit switching, a request is a pair $(\mv a, \mv b)$ where
$\mv a$ is an unused input and $\mv b$ is an unused output.
A route $R(\mv a, \mv b)$ realizes the request if it does not share any
internal link with existing routes.
In an $f$-cast switching network, each multicast request is of
the form $(\mv a, B)$ where $\mv a$ is some input and 
$B$ is a subset of at most $f$ outputs.
The number $f$ is called the {\em fanout restriction}.
An $N\times n$ multicast network without fanout restriction is equivalent to
an $N$-cast network.

In the multirate case, each link has a capacity (e.g., bandwidth).
All inputs and outputs have the same capacity normalized to $1$.
An input cannot request more than its capacity. Neither can outputs.
%All internal links have the same capacity of $\beta \geq 1$.
%If $\beta>1$, then we have {\em internal speedup}, which is a common 
%design technique to increase the total switch capacity \cite{melenturner2003}.
%Internal links have capacity $1$.
A request is of the form $(\mv a, \mv b, w)$ where $\mv a$ is an input,
$\mv b$ is an output, and $w\leq 1$ is the requested {\em rate}.
If existing requests have used up to $x$ and $y$ units of $\mv a$'s  and 
$\mv b$'s capacity, respectively, then the new requested rate
$w$ can only be at most $\min\{1-x,1-y\}$. 
An internal link cannot carry requests with total rate more than $1$. %$\beta$.

%%%%%%%%%%%%%%%%%%%%%%%%%%%%%%%%%%%%%%%%%%%%%%%%%%%%%%%%%%%%%%%%%%%%%%%%%%%%%%%
\subsection{The $3$-stage Clos networks}

\begin{figure}[hbtp]
\centerline{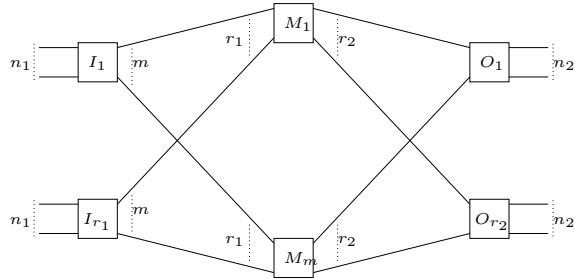}
\caption{The $3$-stage Clos network $C(n_1,r_1,m,n_2,r_2)$}
\label{fig:clos}
\end{figure}

The Clos network $C(n_1,r_1,m,n_2,r_2)$ is a $3$-stage interconnection network,
where the first stage consists of $r_1$ crossbars of size $n_1 \times m$,
the last stage has $r_2$ crossbars of dimension $m \times n_2$,
and the middle stage has $m$ crossbars of dimension $r_1 \times r_2$
(see Figure \ref{fig:clos}).
Each input crossbar $I_i$ ($i=1,\dots,r_1$) is connected to each
middle crossbar $M_j$ ($j=1,\dots,m$).
Similarly, the middle stage and the last stage are fully connected.
When $n_1=n_2=n$ and $r_1=r_2=r$, the network is called the
\emph{symmetric $3$-stage Clos network}, denoted by $C(n,m,r)$.

%%%%%%%%%%%%%%%%%%%%%%%%%%%%%%%%%%%%%%%%%%%%%%%%%%%%%%%%%%%%%%%%%%%%%%%%%%%%%%%
\subsection{The $d$-ary multilog networks}

\begin{figure}[t]
\centerline{\includegraphics[width=2.5in]{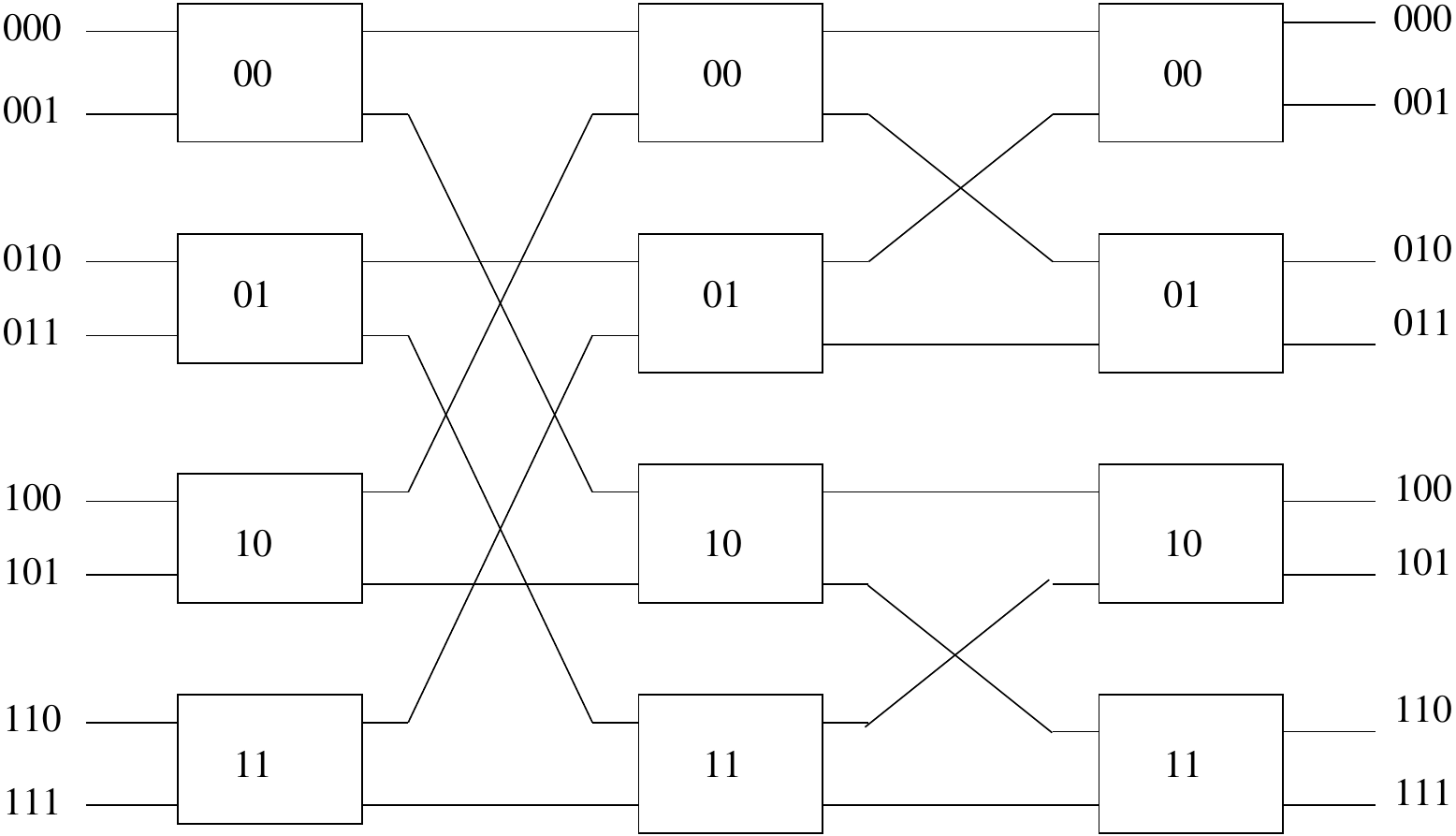}}
\caption{The inverse Banyan network BY$^{-1}(3)$}
\label{fig:BY}
\end{figure}

\begin{figure}[t]
\centerline{\includegraphics[width=2.5in]{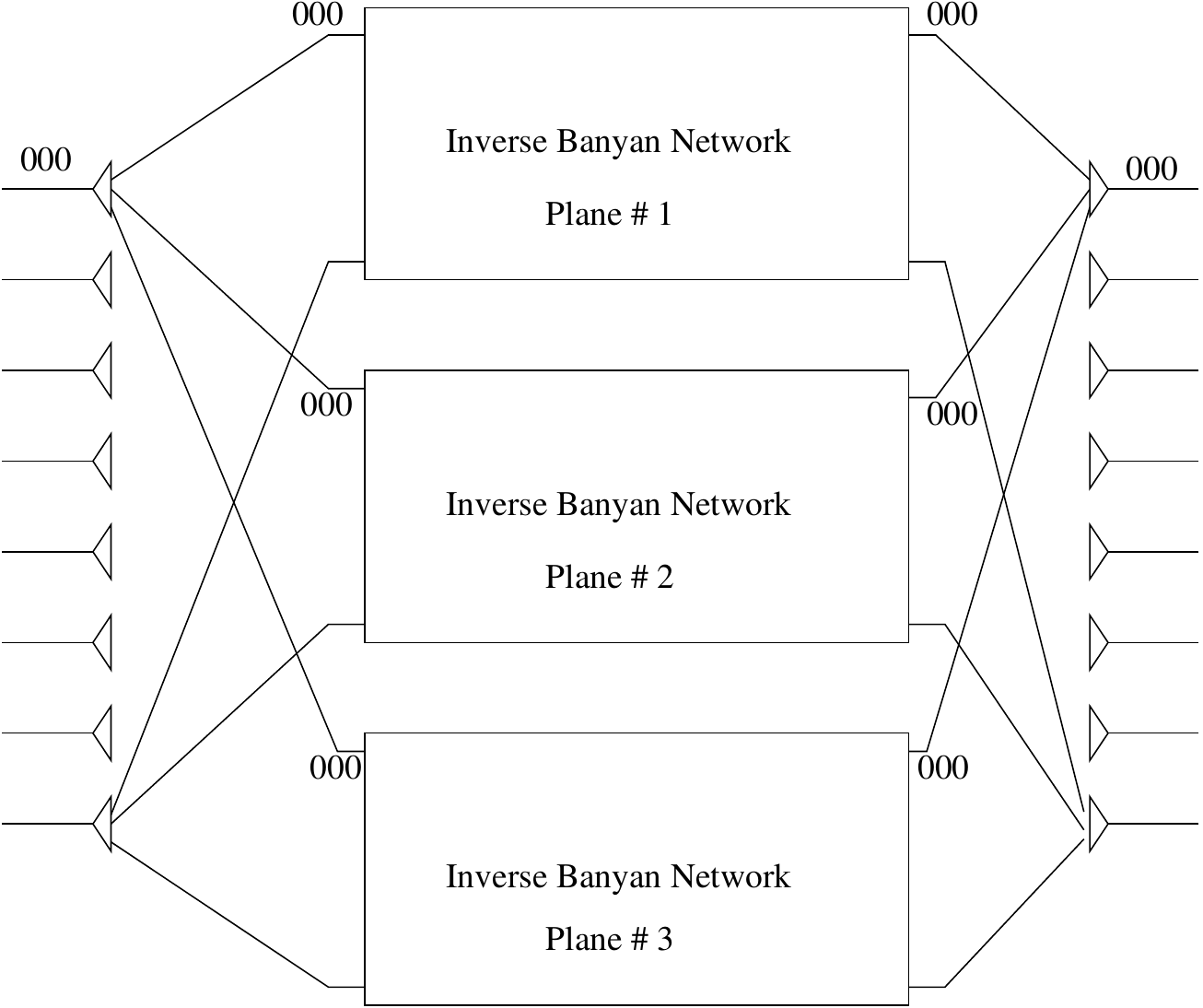}}
\caption{A multi-log network with $3$ inverse Banyan planes}
\label{fig:multilog}
\end{figure}

Let $N=d^n$. We consider the $\log_d(N,0,m)$ network, which denotes
the stacking of $m$ copies of the $d$-ary inverse Banyan network BY$^{-1}(n)$
with $N$ inputs and $N$ outputs.
(See Fig. \ref{fig:BY} and \ref{fig:multilog}.)
Label the inputs and outputs of BY$^{-1}(n)$ 
and the $d\times d$ switching elements (SE) of each stage 
of BY$^{-1}(n)$ as illustrated in Fig. \ref{fig:BY}.
We label the inputs and outputs of a BY$^{-1}(n)$-plane with
$d$-ary strings of length $n$.
Specifically, each input $\mv u \in \mathbb Z_d^n$
and output $\mv v  \in \mathbb Z_d^n$ have the form
$\mv u = u_1\cdots u_n$,
$\mv v = v_1\cdots v_n$,
where $u_i, v_i \in \mathbb Z_d$, $\forall i \in [n]$.
Also, label the $d \times d$ switching elements in
each of the $n$ stages of a BY$^{-1}(n)$-plane with
$d$-ary strings of length $n-1$.
An input $\mv x$ (respectively, output $\mv y$) is
connected to the switching element labeled $\mv x_{1..n-1}$ in the first stage
(respectively, $\mv y_{1..n-1}$ in the last stage).
A switching elements labeled $\mv z = z_1\cdots z_{n-1}$ in stage $i \leq n-1$
is connected to $d$ switching elements in stage $i+1$ numbered
$z_1\cdots z_{i-1} * z_{i+1}\cdots z_{n-1}$, where $*$ is any symbol
in $\mathbb Z_d$.

For the sake of clarity, let us first consider a small example.
Consider the unicast request $(\mv x, \mv y) = (01001, 10101)$
when $d=2, n=5$.
The input $\mv x = 01001$ is connected to the switching element labeled $0100$ in
the first stage, which
is connected to two switching elements labeled $0100$ and $1100$ in the
second stage, and so on.
The unique path from $\mv x$ to $\mv y$ in the BY$^{-1}(n)$-plane
can be explicitly written out (see Figure \ref{fig:BY2}):
\begin{figure}[t]
\centerline{\includegraphics[width=2.6in]{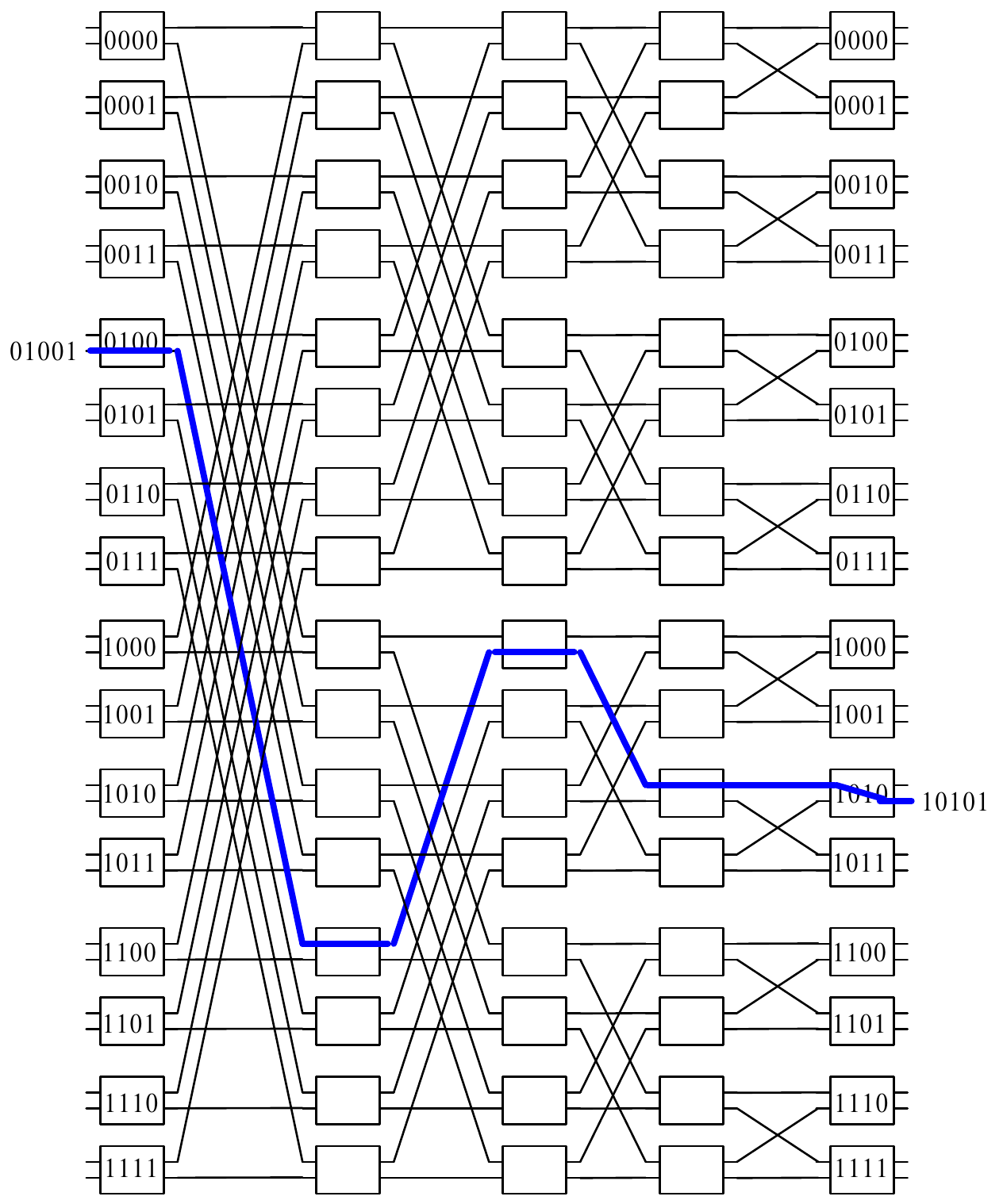}}
\caption{The inverse Banyan network BY$^{-1}(5)$}
\label{fig:BY2}
\end{figure}
\begin{center}
\begin{tabular}{l|l}
input $\mv x$ & $01001$ \\
\hline
stage-$1$ switching element & $0100$ \\
stage-$2$ switching element & $\mv{1}100$ \\
stage-$3$ switching element & $\mv{10}10$ \\
stage-$4$ switching element & $\mv{101}0$ \\
stage-$5$ switching element & $\mv{1010}$ \\
\hline
output $\mv y$ & $10101$ \\
\end{tabular}
\end{center}
We can see clearly the pattern: the prefixes of $\mv y_{1..n-1}$ are
``taking over'' the prefixes of $\mv x_{1..n-1}$ on the path from $\mv x$ to
$\mv y$.
In general, the unique path $R(\mv x, \mv y)$ in
a BY$^{-1}(n)$-plane from an arbitrary input $\mv x$ to an arbitrary
output $\mv y$ is exactly the following:
\begin{center}
\begin{tabular}{c|l}
input $\mv x$ & $x_1x_2\dots x_{n-1}x_n$ \\
\hline
stage-$1$ switching element & $x_1x_2\dots x_{n-1}$ \\
stage-$2$ switching element & $y_1x_2\dots x_{n-1}$ \\
stage-$3$ switching element & $y_1y_2\dots x_{n-1}$ \\
$\vdots$ & $\vdots$ \\
stage-$n$ switching element & $y_1y_2\dots y_{n-1}$ \\
\hline
output $\mv y$ & $y_1y_2\dots y_{n-1}y_n$ \\
\end{tabular}
\end{center}

%It is easy to see that
%an input $\mv u$ (resp. output $\mv v$) is
%connected to the SE labeled $\mv u_{1..n-1}$ in the first stage
%(resp. $\mv v_{1..n-1}$ in the last stage).
%In general, the unique path $R(\mv u, \mv v)$ in
%BY$^{-1}(n)$ from an arbitrary input $\mv u$ to an arbitrary
%output $\mv v$ is exactly the following:
%input $\mv u  = u_1u_2\dots u_{n-1}u_n$ to
%stage-$1$ SE $u_1u_2\dots u_{n-1}$ to
%stage-$2$ SE $v_1u_2\dots u_{n-1}$ to
%stage-$3$ SE $v_1v_2\dots u_{n-1}$, $\dots$,
%to stage-$n$ SE $v_1v_2\dots v_{n-1}$, to
%output $\mv v$ $v_1v_2\dots v_{n-1}v_n$.

%\begin{figure}[t]
%\centerline{\includegraphics[width=2.6in]{log3(27,2,0)}}
%\caption{The $\log_3(27,0,2)$ network}
%\label{fig:multilog}
%\end{figure}

Now, consider two unicast requests
$(\mv a, \mv b)$ and $(\mv x, \mv y)$.
From the observation above,
on the same BY$^{-1}(n)$-plane
the two routes $R(\mv a, \mv b)$ and $R(\mv x, \mv y)$
share a switching element (also called a node)
if and only if there is some $j\in [n]$ such that
$b_{1..j-1} = y_{1..j-1}$ and $a_{j..n-1} = x_{j..n-1}$.
In this case, the two paths intersect at a stage-$j$ switching element.
It should be noted that two requests' paths may intersect at more
than one switching element.

For any two $d$-ary strings $\mv u, \mv v \in \mathbb Z_d^l$,
let $\prefix(\mv u, \mv v)$ denote the {\em longest common prefix},
and $\suffix(\mv u, \mv v)$ denote the {\em longest common suffix} of
$\mv u$ and $\mv v$, respectively.
For example, if $\mv u = 0100110$ and $\mv v = 0101010$, then
$\prefix(\mv u, \mv v) = 010$ and $\suffix(\mv u, \mv v) = 10$.
The following propositions straightforwardly follow
(for more details, see e.g. \cite{wang-icc2008}).

\begin{prop}
Let $(\mv a, \mv b)$ and $(\mv u, \mv v)$ be two unicast
requests. Then their corresponding routes
$R(\mv a, \mv b)$ and $R(\mv u, \mv v)$ in
a BY$^{-1}(n)$-plane
share at least a common SE if and only if
\beq |\suffix(\mv a_{1..n-1}, \mv u_{1..n-1})| +
     |\prefix(\mv b_{1..n-1}, \mv v_{1..n-1})| \geq n-1.
\eeq
Moreover, the routes $R(\mv a, \mv b)$ and $R(\mv u, \mv v)$
intersect at exactly one SE if and only if
\beq |\suffix(\mv a_{1..n-1}, \mv u_{1..n-1})| +
     |\prefix(\mv b_{1..n-1}, \mv v_{1..n-1})| = n-1,
\eeq
in which case the common SE is an SE at
stage
$|\prefix(\mv b_{1..n-1}, \mv v_{1..n-1})|+1$ of
the BY$^{-1}(n)$-plane.
\label{prop:unicast-node-block}
\end{prop}

\begin{prop}
Let $(\mv a, \mv b)$ and $(\mv u, \mv v)$ be two unicast
requests. Then their corresponding routes
$R(\mv a, \mv b)$ and $R(\mv u, \mv v)$ in
a BY$^{-1}(n)$-plane
share at least a common link iff
\beq |\suffix(\mv a_{1..n-1}, \mv u_{1..n-1})| +
     |\prefix(\mv b_{1..n-1}, \mv v_{1..n-1})| \geq n.
\eeq
\label{prop:unicast-link-block}
\end{prop}
%\input{clos}

%%%%%%%%%%%%%%%%%%%%%%%%%%%%%%%%%%%%%%%%%%%%%%%%%%%%%%%%%%%%%%%%%%%%%%%%%%%%%%%
\section{Results on the Clos Networks}
\label{sec:clos}
%%%%%%%%%%%%%%%%%%%%%%%%%%%%%%%%%%%%%%%%%%%%%%%%%%%%%%%%%%%%%%%%%%%%%%%%%%%%%%%

%%%%%%%%%%%%%%%%%%%%%%%%%%%%%%%%%%%%%%%%%%%%%%%%%%%%%%%%%%%%%%%%%%%%%%%%%%%%%%%
\subsection{Two classic examples in circuit switching}
%\subsection{A classic example in circuit switching}

%\iffalse
To illustrate the LP-duality technique, we begin with two 
simple examples which have become classic textbook materials.

\begin{ex}[The SNB Case]
Consider the symmetric Clos network $C(n,m,r)$.
Consider a new request from an input of input crossbar $I$ to an output
of output crossbar $O$. 
A middle crossbar cannot carry this request if it already carried
some request from $I$ or some request to $O$.
Let $x$ (resp. $y$) be the number of middle crossbars
which already carry some requests from $I$ (resp. to $O$).
Since the number of existing requests from $I$ or to $O$ is at most
$n-1$, we have $x\leq n-1$ and $y\leq n-1$. The number of unavailable
middle crossbars is thus bounded above by the
optimal value of the LP
\[ \max\{x+y \suchthat x\leq n-1,y\leq n-1, x,y\geq 0\}. \]
The dual program is
\[ \min\{(n-1)(\alpha+\beta) \suchthat \alpha \geq 1, \beta \geq 1, 
   \alpha,\beta \geq 0\}. 
\]
Setting $\alpha=\beta=1$ is certainly dual-feasible, and thus
its objective value $2n-2$ is an upper bound on the number of
unavailable middle crossbars. We conclude that
$m\geq 2n-1$ is sufficient for $C(n,m,r)$ to be SNB.
\label{ex:1}
\end{ex}
%\fi

\begin{ex}[The WSNB Case]
This example is a classic result by Benes \cite{MR35:1408}.
Consider the $C(n,m,2)$ network.
The routing algorithm is simply the following rule: {\em reuse a busy middle 
crossbar whenever possible.}

For any $i,j\in\{1,2\}$, let
$M_{ij}$ be the set of middle crossbars carrying
an $I_i,O_j$-request.
The sets $M_{ij}$ certainly change over time as requests come and go.
However, it is easy to show by induction that the routing rule ensures
$|M_{11}\cup M_{22}|\leq n$ and
$|M_{12}\cup M_{21}|\leq n$ at all times.
To see this, without loss of generality consider a new $I_1, O_1$-request.
If we can find a crossbar in $M_{22}$ to route the new request, then
the union $M_{11}\cup M_{22}$ does not change and thus 
$|M_{11}\cup M_{22}|\leq n$ by induction hypothesis.
If every crossbar in $M_{22}$ is not available for the new request, 
then it must be the case that $M_{22} \subseteq M_{11}$.
There are at most $n-1$ existing requests out of $I_1$. Thus, 
$|M_{11}| \leq n-1$. Hence, before routing the new $I_1, O_1$-request
we have $|M_{11}\cup M_{22}| = |M_{11}| \leq n-1$. Consequently, after
realizing the new request, we have $|M_{11}\cup M_{22}|\leq n$.

Next, again without loss of generality,
consider a new request from $I_1$ to $O_1$.
If $M_{22} \setminus M_{11} \neq \emptyset$, then we have a busy crossbar
to reuse.
Otherwise, 
the number of unavailable middle-crossbars for this new request
is precisely $|M_{11}\cup M_{12} \cup M_{21}| = |M_{11}|+|M_{12}\cup M_{21}|$.
Just before the arrival of this new request, 
the number of existing requests {\em from} $I_1$ or {\em to} $O_1$ is 
at most $n-1$,
i.e. $|M_{11}\cup M_{12}| = |M_{11}|+|M_{12}| \leq n-1$,
and $|M_{11}\cup M_{21}| = |M_{11}|+|M_{21}| \leq n-1$.
The number of {\em unavailable} middle crossbars is thus bounded by the 
optimal
value of the following LP, where we think of set cardinalities as variables:
\begin{eqnarray*}
 \max & |M_{11}|+|M_{12}\cup M_{21}| &\\
 \sto & |M_{11}|+|M_{12}| & \leq n-1 \\
      & |M_{11}|+|M_{21}| & \leq n-1 \\
      & |M_{12}|+|M_{21}| & \leq n \\
      & |M_{12}\cup M_{21}|-|M_{12}|-|M_{21}| & \leq 0
\end{eqnarray*}
The last inequality is the straightforward union bound. Obviously,
all cardinalities are non-negative. The dual LP is
\begin{eqnarray*}
 \min & (n-1)(y_1+y_2) + ny_3 &\\
 \sto & y_1+y_2 & \geq 1 \\
      & y_2+y_3-y_4 & \geq 0 \\
      & y_1+y_3-y_4 & \geq 0 \\
      & y_4 \geq 1, \ \ y_1,y_2,y_3& \geq 0
\end{eqnarray*}
Setting $y_1=y_2=y_3=1/2$ and $y_4=1$ is certainly dual-feasible
with objective value $3n/2-1$. Hence, by weak duality
the number of unavailable
middle-crossbars for the new $I_1,O_1$-request is at most 
$\lfloor 3n/2 \rfloor-1$,
which means $m\geq \lfloor 3n/2\rfloor$ is sufficient
for $C(n,m,2)$ to be WSNB.
It is not hard to show that $m \geq \lfloor 3n/2 \rfloor$ is also necessary 
\cite{MR35:1408}.
This ($r=2$) is the only case for which a necessary and sufficient condition
is known for the Clos network $C(n,m,r)$ to be WSNB!
\end{ex}

%%%%%%%%%%%%%%%%%%%%%%%%%%%%%%%%%%%%%%%%%%%%%%%%%%%%%%%%%%%%%%%%%%%%%%%%%%%%%%%
\subsection{Multirate switching and the {\sc dwec} problem}

It is known that $C(n,m,r)$
is multirate WSNB when $m\geq 5.75n$ \cite{MR98e:68008}. 
This section uses the LP technique to improve this bound via solving
a much more general problem called {\sc dynamic weighted edge coloring}
({\sc dwec}).

\bdefn[The {\sc dwec} problem]
Let $G=(V,E)$ be a fixed simple graph called the {\em base graph}.
% and $\beta\geq 1$ be a given parameter.
Let $G_0=(V,\emptyset)$ be an empty graph with the same vertex set.
At time $t$, either an arbitrary edge $e$ is removed from $G_{t-1}$,
in which case $G_t = G_{t-1}-\{e\}$,
or a copy of some edge $e\in E$ ``arrives'' along with a weight
$w_e \in (0,1]$, in which case define $G_t = G_{t-1} \cup\{e\}$.
Note that $G_t$ can be a multi-graph as many copies of the same edge
may arrive over time.
The arriving edge is to be colored so that, in $G_t$,
the total weight of same-color edges incident to the same vertex
is at most $1$.

The objective is to design a coloring algorithm so that the number
of colors used is minimized, compared to an off-line algorithm which
colors edges of $G_t$ subject to the same constraint.
Formally, let $\opt(t)$ denote the number of colors used
by an optimal off-line algorithm on $G_t$.
Let $\overline \opt(t) = \max_{i\leq t} \opt(i)$.
For any online coloring algorithm $\mv A$, let
$\bar{\mv A}(t)$ be the number of colors ever used by $\mv A$  up to time $t$.
Algorithm $\mv A$ has {\em competitive ratio} $\rho$ if,
for any sequence of edge arrivals/departures with arbitrary weights,
we always have $\bar{\mv A}(t) \leq \rho \cdot \overline \opt(t), \forall t$.
\edefn

The {\sc dynamic bin packing} problem is {\em exactly} the {\sc dwec} problem
when the base graph $G = K_2$, where each color is a bin.
The best competitive ratio for
{\sc dynamic binpacking} is known
to be between $2.5$ and $2.788$ \cite{MR697157}.
We will show that the {\sc dwec}'s best competitive ratio is
somewhere between $4$ and $5.6355$ for any base graph $G$.

%The following theorem is the only example in the paper which does not need
%to involve the dual program; partly because the primal program is 
%small enough to the point that we can solve it directly.
\bthm
%When $\beta=1$,
There is an algorithm for {\sc dwec} with competitive ratio $5.6355$.
\ethm
\bp
For the sake of presentation clarity, we will prove a slightly weaker 
ratio of $5.675$, and then
indicate how to obtain the better ratio $5.6355$.
The two proofs are identical, but the one we present is cleaner.

At any time $t$, let $W^u(t)$ denote the total weight 
of edges incident to $u$ in $G_t$,
and let $d^u(t)$ denote the number of edges of weight $>1/2$
incident to $u$.
Let $\overline W(t) = \max_{i \leq t} \max_u W^u(i)$ and 
$\overline \Delta(t) = \max_{i \leq t}  \max_ud^u(i)$.
It is not hard to see that
$\lceil \overline W(t) \rceil \leq \overline \opt(t)$
and $\overline \Delta(t) \leq \overline \opt(t)$.

Refer to an edge a
{\em type-$0$}, {\em type-$1$},
{\em type-$2$}, or {\em type-$3$},
if its weight belongs to the interval
$(\frac 1 2, 1]$,
$(\frac 2 5, \frac 1 2]$,
$(\frac 1 3, \frac 2 5]$, or
$(0, \frac 1 3]$, respectively.
Our coloring algorithm is as follows.
Maintain $4$ {\em disjoint} sets of colors $C_i(t)$, $0\leq i \leq 3$.
Let $x_0,x_1,x_2,x_3$ be constants to be determined.
For each $i=0..3$, we will maintain the following
time-invariant conditions: $|C_i(t)| = \lceil x_i \overline{W}(t)\rceil$ for
$1\leq i\leq 3$ and $|C_0(t)| = \lceil x_0 \overline\Delta(t) \rceil$.

If $\overline W(t)$ or $\overline \Delta(t)$
is increased at some time $t$ then we are allowed
to add new colors to the sets $C_i(t)$ to maintain the invariants.
Note that $\overline W(t)$ and $\overline\Delta(t)$ are non-decreasing in $t$; 
hence, colors will never be removed from the $C_i(t)$.
The colors in $C_0(t)$ are used exclusively for edges of type-$0$.
The coloring for edges of types $i$, $1\leq i\leq 3$ is done as follows.
If a type-$i$ edge arrives at time $t$, find a color in $C_i(t)$
to color it. If $C_i(t)$ cannot accommodate this edge, try $C_{i+1}(t)$,
and so on until $C_3(t)$.
We next show that if the constants $x_i$ are feasible solutions to
a certain LP, then it is always possible to color an arriving edge. 
%To simplify notations, define $c_i = |C_i(t)|$.

Suppose a type-$0$ edge $e=(u,v)$ arrives at time $t$.
If we cannot find a color in $C_0(t)$ for $e$, then
$|C_0(t)| \leq d^u(t-1) +d^v(t-1) = (d^u(t)-1)+(d^v(t)-1) <
2\overline\Delta(t)$. Hence, as long as $x_0\geq 2$ we can color $e$.

Next, suppose $e=(u,v)$ of type $1$ arrives at time $t$
and we cannot find a color in $C_1(t) \cup C_2(t) \cup C_3(t)$ to color $e$.
For a color $c \in C_1(t)$ to be unavailable for $e$, there must be at least
two type-$1$ color-$c$ edges incident to either $u$ or $v$.
Thus, the total type-$1$ weight at $u$ and $v$ is 
$> \frac 4 5 |C_1(t)|$.
Similarly,
for each color $c$ in $C_2(t)$, the total $c$-weight incident to $u$ and $v$
must be $> 1/2$, which means this color $c$ ``carries''
either at least two type-$1$ edges,
or one type-$1$ edge and one type $2$ edge,
or at least two type-$2$ edges.
Thus, the total color-$c$ weight incident to $u$ and $v$ must be
$> \frac 2 3 |C_2(t)|$.
Lastly, for each color $c$ in $C_3(t)$, the total color-$c$ weight
incident to $u$ and $v$ must be $>1/2 |C_3(t)|$.
Note that the total weight at $u$ and $v$ is $<2\overline W(t)$.
Consequently, we will be able to find a color for $e$ if
\[ \frac 4 5|C_1(t)| + \frac 2 3|C_2(t)| + \frac 1 2|C_3(t)|
   \geq 2 \overline W(t),
\]
which would hold if
$\frac 4 5 x_1 + \frac 2 3 x_2 + \frac 1 2 x_3 \geq 2.$
Similarly, a newly arriving type-$2$ edge is colorable if
$\frac 2 3 x_2 + \frac 3 5 x_3 \geq 2,$
and a new type-$3$ edge is colorable if $\frac 2 3 x_3 \geq 2$.
Consequently, our coloring algorithm works if the $x_i$
are feasible for the following LP:
\[
\begin{matrix}
\min & x_0&+x_1&+x_2&+x_3 \\
\sto & x_0 & & & & \geq & 2\\
 &     & \frac 4 5 x_1 & + \frac 2 3 x_2 & + \frac 1 2 x_3 & \geq & 2\\
 &     & & \frac 2 3 x_2 & + \frac 3 5 x_3 & \geq & 2\\
 &     & & & \frac 2 3 x_3 & \geq & 2\\
 & \multicolumn{4}{r}{x_0,x_1,x_2,x_3} & \geq & 0.
\end{matrix}
\]
The solution $x_0=2, x_1=3/8, x_2=3/10, x_3=3$ is certainly
feasible. The total number of colors used is
\[
 \lceil x_0\overline\Delta(t) \rceil +
   \sum_{i=1}^3 \lceil x_i\overline W(t) \rceil 
 \leq (x_0+x_1+x_2+x_3)\overline\opt(t) + \frac 7 8 + \frac{9}{10}
  \leq 5.675\overline\opt(t)+1.8.
\]
As is customary in online/dynamic algorithm analysis, we ignore the
constant term of $1.8$, as we let $\overline\opt(t)  \to \infty$.
To prove the better ratio $5.6355$,
divide the rates into $5$ types belonging to the
intervals $(1/2,1]$, $(2/5,1/2]$, $(1/3,2/5]$, $(11/43,1/3]$, and $(0,11/43]$.
\ep

\bcor
The Clos network $C(n,m,r)$ is multirate WSNB if
$m \geq 5.6355n+4$.
\ecor
\bp
Consider the multirate WSNB problem on the Clos network $C(n,m,r)$. 
We formulate a {\sc dwec} instance generalizing the problem.
The base graph is the complete bipartite graph 
$G= \mathcal I \times \mathcal O$,
where $\mathcal I$ is the set of input crossbars and $\mathcal O$ is the set 
of output crossbars.
When a new request $(\mv a, \mv b, w)$ arrives at time $t$, 
add an edge $e=(I,O)$ to $G_{t-1}$ where $I$ is the input crossbar 
to which $\mv a$ belongs and $O$ is the output crossbar 
to which $\mv b$ belongs.
Set the edge weight $w_e=w$.
Think of each middle-crossbar as a color.
Obviously, the maximum number of colors ever used by an algorithm $\mv A$
is also a sufficient number of middle crossbars needed for 
$C(n,m,r)$ to be non-blocking.

In the above algorithm, $\overline\Delta(t) \leq n$ because the number
of requests with rate $>1/2$ coming out of the same input crossbar or into
the same output crossbar is at most $n$ (one per input/output).
Moreover, $\overline W(t) \leq n$ because the total rate of requests from/to
an input/output is at most $n$. 
Hence, the number of middle-stage crossbars (i.e. colors) needed is
at most $5.6355n+4$.
\ep

\brmk
Our strategy can also give a better sufficient condition
than the best known in \cite{MR98e:68008} for the case
when there's internal speedup in the Clos network.
However, for the ease of exposition,
we refrain from stating the most general result we can prove.
\ermk

\section{Analyzing $f$-cast wide-sense nonblocking multilog networks}
\label{sec:fcast-WSNB}

Let $f,t$ be given integers with $0\leq t \leq n$,
and $1\leq f \leq N=d^n$.
This section analyzes $f$-cast wide-sense nonblocking $\log_d(N,0,m)$ networks
under the {\em window algorithm} with window size $d^t$.
The algorithm was proposed and analyzed for one window size 
$d^{\lfloor n/2 \rfloor}$ in \cite{Tscha-Lee}, and 
later analyzed more carefully for varying window sizes in 
\cite{DBLP:journals/tcom/Danilewicz07}. 
Both papers considered the multicast case with no fanout restriction.
We will derive a more general theorem for the $f$-cast case.
\bi
 \item {\bf The Window Algorithm with window size $d^t$:} 
Given any integer $t$, $0\leq t\leq n$, divide the outputs into
``windows'' of size $d^t$ each. 
Each window consists of all outputs sharing
a prefix of length $n-t$, for a total of $d^{n-t}$ windows.
Denote the windows by $W_w,0\leq w\leq d^{n-t}-1$.
Given a new multicast request $(\mv a, B)$, where $\mv a$ is an input
and $B$ is a subset of outputs,
the routing rule is, for every $0\leq w \leq d^{n-t}-1$,
the subrequest $(\mv a, B \cap W_w)$ is routed entirely on 
one single BY$^{-1}(n)$-plane.
(Different sub-requests can be routed through the same or different
BY$^{-1}(n)$-planes.)
\ei

\brmk there is a subtle point about the window algorithm
due to which the original authors in \cite{Tscha-Lee} thought their
multilog network was strictly nonblocking instead of wide-sense nonblocking. Basically, for some specific values
of the parameters the algorithm is {\bf no} algorithm at all.
In those cases, any sufficient condition for the network to be nonblocking
under the window algorithm is in fact a strictly nonblocking condition, 
not a wide-sense nonblocking condition.

%Obvious examples include the unicast ($f=1$) case and the $t=0$ case.
For example, in the unicast case we have $f=1$, which means
the window algorithm does not specify
any routing strategy; consequently, any nonblocking condition is actually
a strictly non-blocking condition.
Another example is when $t=0$. In this case, the routing rule says that
each branch of a (multicast) request should be routed on some plane, 
independent of other branches. 
Because there is no restriction on how to route the branches,
any nonblocking condition is a strictly non-blocking one.

Yet another example is when $t=n$. Here, the routing rule
is for each request to be routed entirely on  some plane.
If the $1 \times m$-SE stage of the multilog network has
fanout capability, then the rule {\bf does} restrict how we route requests,
and thus we indeed have a wide-sense nonblocking situation.
However, if the $1 \times m$-SE stage is implemented with
$1 \times m$-unicast crossbars or $1 \times m$-demultiplexers, then 
we {\em have to} route each request entirely on some plane. 
Thus, any sufficient condition is a strictly nonblocking condition.
\label{rmk:snb-vs-wsnb-remark}
\ermk

\subsection{Setting up the linear program and its dual}

Let $(\mv a, B)$ be an arbitrary $f$-cast request to be routed
using the window algorithm with window size $d^t$.
Following the window algorithm, due to symmetry
without loss of generality we can assume that
$B = \{\mv b^{(1)}, \dots, \mv b^{(k)}\}$ where
all the outputs $\mv b^{(l)}$ ($l\in [k]$) belong to the same window $W_0$,
and $k\leq \min\{f,d^t\}$.
The $\mv b^{(l)}$ thus share a common prefix of length $n-t$.
(This is because subrequests to the same window are routed through the same
plane and different subrequests of the same request are routed independently
from each other and they do not block one another.)

For each $i \in \{0,\dots,n-1\}$, let
$A_i$ be the set of inputs $\mv u$ other than ${\mv a}$, where
$\mv u_{1..n-1}$ shares a {\em suffix} of length exactly $i$ with
${\mv a}_{1..n-1}$. Formally, define
%Define
\[ A_i := \left\{ \mv u \in \mathbb Z_d^n - \{\mv a\} \suchthat
                   \suffix(\mv u_{1..n-1}, {\mv a}_{1..n-1}) = i
          \right\}.
\]
For each $j \in \{0,\dots,n-1\}$, let $B_j$ be the set
of outputs other than those in $B$ which
share a {\em prefix} of length exactly $j$ with
{\em some} member of $B$, namely
\[ B_j := \left\{ \mv v \in \mathbb Z_d^n - B
                   \suchthat \exists l \in [k],
                   \prefix(\mv v_{1..n-1}, {\mv b}^{(l)}_{1..n-1}) = j
          \right\}.
\]
Note that 
\begin{eqnarray*}
|A_i| &=& d^{n-i}-d^{n-1-i}, 0\leq i \leq n-1,\\
|B_j| &=& d^{n-j}-d^{n-1-j}, 0\leq j \leq n-t-1.
\end{eqnarray*}
Define $\mathcal A = \bigcup_{i=0}^{n-1} A_i$.
%$\mathcal B = \bigcup_{j=0}^{n-t-1} B_j$.
It is easy to see that
\begin{eqnarray*}
\bigcup_{j=0}^{n-t-1} B_j &=& \bigcup_{w=1}^{d^{n-t}-1} W_w\\
\bigcup_{j=n-t}^{n-1} B_j &=& W_0 - B.\\
\end{eqnarray*}
Furthermore, for each $j\leq n-t-1$, $B_j$ is the disjoint union
of precisely $d^{n-j-t}-d^{n-1-j-t}$ windows each of size $d^t$.

Note that the sets $B_j$ for $0\leq j \leq n-t-1$ are mutually disjoint.
On the other hand, the sets $B_j$ for $n-t\leq j \leq n-1$ are not necessarily 
disjoint, 
because for the same output $\mv v \in W_0-B$ it might be the case
that $\prefix(\mv v_{1..n-1}, {\mv b}^{(l)}_{1..n-1}) = j$
and $\prefix(\mv v_{1..n-1}, {\mv b}^{(l')}_{1..n-1}) = j'$
for $j\neq j'$, $l\neq l'$.
The following simple observation turns out to be an important analytical
detail in many of the proofs.

\bprop
Let $q$ be an integer such that $n-t\leq q \leq n-1$. Then,
\[ \left|\bigcup_{j=q}^{n-1} B_j\right| \leq \min\{d^t-k, k(d^{n-q}-1)\},
\]
and
\[ \left|\bigcup_{j=n-t}^{n-1} B_j\right| = d^t-k. \]
\label{prop:q-bound}
\eprop
\bp
To see the inequality, note that 
$\left|\bigcup_{j=q}^{n-1} B_j\right|$ counts the number of strings $\mv v$
in $W_0-B$ for which 
\[ \prefix(\mv v_{1..n-1}, {\mv b}^{(l)}_{1..n-1}) \geq q \]
for some $\mv b^{(l)}, l \in [k]$.
As $|W_0|=d^t$, the upper-bound $d^t-k$ for the number of such strings is
trivial.
On the other hand, the number of strings $\mv v$ where
$\prefix(\mv v_{1..n-1}, {\mv b}^{(l)}_{1..n-1}) \geq q$
for a fixed string $\mv b^{(l)}$ is at most
$d^{n-q}-1$ (discounting $\mv b^{(l)}$ itself).
Hence, we get the upper-bound $k(d^{n-q}-1)$ via a simple application of
the union bound.
The equality trivially holds because
$\bigcup_{j=n-t}^{n-1} B_j = W_0 - B$.
\ep

For every input $\mv u\in \mathcal A$, let $i(\mv u)$ denote
the index $i$ such that $\mv u \in A_i$.
For every $w\in [d^{n-t}-1]=\{1,\dots,d^{n-t}-1\}$, 
let $j(w)$ be the index $j$ such that $W_w \subseteq B_j$.
For every $\mathbf v \in W_0-B$, let $j(\mathbf v)$ denote the
{\em largest} $j$ for which $\mathbf v \in B_j$. Note that
$j(\mathbf v)\geq n-t$ for such output $\mathbf v$
because $W_0-B = \bigcup_{j=n-t}^{n-1} B_j$.

\blmm
For each input $\mv u \in \mathcal A$ and
each $w\in [d^{n-t}-1]$ such that $i(\mv u)+j(w)\geq n$,
define a variable $x_{\mv u,w}$.
Also,
for each input $\mv u \in \mathcal A$ and
each output $\mathbf v\in W_0-B$ such that
$i(\mv u)+j(\mv v) \geq n$,
define a variable $x_{\mv u,\mv v}$.
Then, the number of Banyan planes blocking the new multicast request
$(\mv a, B)$ is upperbounded
by the optimal value of the following linear program:
\begin{equation}
\begin{matrix}
\max & \multicolumn{4}{l}{\displaystyle{\sum_{\mv u, w} x_{\mv u,w} + 
       \sum_{\mv u, \mv v} x_{\mv u,\mv v}}}\\
\sto & \sum_{\mv u} x_{\mv u,w} & \leq & d^t & w \in [d^{n-t}-1]\\
     & x_{\mv u,w} & \leq & 1 & \forall \mv u, w\\
     & \sum_{\mv v} x_{\mv u, \mv v} & \leq & 1 & \forall \mv u \in \mathcal A\\
     & \sum_{\mv u} x_{\mv u,\mv v}  & \leq & 1 & \forall \mv v \in W_0-B \\
     & \sum_w x_{\mv u,w} + \sum_{\mv v} x_{\mv u,\mv v}  & \leq & f & 
        \forall \mv u \in \mathcal A \\
     & x_{\mv u,w}, x_{\mv u,\mv v} & \geq & \mv 0 & \forall \mv u, w,\mv v
\end{matrix}
\label{eqn:LP}
\end{equation}
Obviously, the sums and the constraints only range over values for which 
the variables are defined.
\label{lmm:LP}.
\elmm
\bp
Suppose the network $\log_d(N,0,m)$ already had some routes established.
Consider a BY$^{-1}(n)$-plane which blocks the new request $(\mv a, B)$.
There must be one route $R(\mv u, \mv v)$ on this plane
for which $R(\mv u, \mv v)$ and $R(\mv a, \mv b^{(l)})$ share a link,
for some $l\in [k]$. Note that the branch $R(\mv u, \mv v)$ could be part
of a multicast tree from input $\mv u$, but we only need an arbitrary
blocking branch $(\mv u, \mv v)$ of this tree.
Note also that $\mv u \neq \mv a$ because subrequests from the same
input are parts of the same request and thus their routes do not
block one another.
Let $S$ be the set constructed by arbitrarily taking exactly one blocking
branch $(\mv u, \mv v)$ per blocking plane.
Then, the number of blocking planes is $|S|$.

{\bf Fact 1:} if $(\mv u, \mv v)$ and $(\mv u, \mv v')$ are both in $S$
then $\mv v$ and $\mv v'$ must belong to different windows; because,
if they belong to the same window, the window algorithm would have routed
them through the same plane, and $S$ only contains one branch per blocking 
plane. 

{\bf Fact 2:} each output $\mv v$ can only appear once in $S$, because
each output can only be part of at most one existing request.

{\bf Fact 3:} if $(\mv u,\mv v)\in S$, then $(\mv u,\mv v) \in A_i\times B_j$
for some $i+j\geq n$, thanks to Proposition \ref{prop:unicast-link-block}.

Straightforwardly, we will show that $S$ defines a feasible solution to the 
linear program
with objective value precisely $|S|$.
Set $x_{\mv u,w}=1$ if there is some $(\mv u,\mv v)\in S$
such that $\mv v\in W_w$;
and $x_{\mv u,\mv v}=1$ if there is some $(\mv u,\mv v) \in S$
such that $\mv v\in W_0-B$.
All other variables are set to $0$.
Due to Fact 3, the procedure does not set
value for an undefined variable. 
Certainly $|S|$ is equal to the objective value of this solution.

We next verify that  the solution satisfies all the constraints.
The first constraint expresses the fact that each output in a window
$W_w$ of size $d^t$ only appears at most once in $S$ (Fact 2).
The second and third constraints are a restatement of Fact 1.
Note that the sumin the third constraint is only over $\mv v\in W_0-B$.
The fourth constraint says that each output $\mv v \in W_0-B$ appears
at most once in $S$ (Fact 2 again).
The fifth constraint says that each input can only be part of
at most $f$ members of $S$, due to the $f$-cast nature of the network.
\ep
The dual linear program can be written as follows.
\begin{equation}
\begin{matrix}
\min & \multicolumn{4}{l}{\displaystyle{\sum_{w} d^t \alpha_{w} +
       \sum_{\mv u, w} \beta_{\mv u,w}} +
       \sum_{\mv u} \gamma_{\mv u} +
       \sum_{\mv v} \delta_{\mv v} +
       \sum_{\mv u} f \epsilon_{\mv u}}\\
\sto 
& \alpha_w + \beta_{\mv u,w} + \epsilon_{\mv u} & \geq & 1, & 
  x_{\mv u, w} \text{ defined} \ \ \textnormal{(DC-1)}\\
& \gamma_{\mv u} + \delta_{\mv v} + \epsilon_{\mv u} & \geq & 1, &
  x_{\mv u, \mv v} \text{ defined} \ \ \textnormal{(DC-2)}\\
& \alpha_w, \beta_{\mv u,w}, \gamma_{\mv u}, \delta_{\mv v}, \epsilon_{\mv u}
 & \geq & 0 & \forall \mv u,\mv v,w
\end{matrix}
\label{eqn:dual-LP}
\end{equation}
Note that the dual-constraints only exist
over all $\mv u,\mv v,w$ for which
$x_{\mv u,w}$ and $x_{\mv u,\mv v}$ are defined, in particular
they exist for pairs $(\mv u, w)$ such that $i(\mv u)+j(w)\geq n$
and pairs $(\mv u, \mv v)$ such that $i(\mv u) + j(\mv v) \geq n$.

\subsection{Specifying a family of dual-feasible solutions}

To illustrate the technique, let us first derive a couple of known results
``for free.'' The first is Theorem III.2 in \cite{wang-icc2008}. 
\bcor[Theorem III.2 in \cite{wang-icc2008}]
Let $r = \lfloor \log_d f \rfloor$.
Suppose the $1\times m$-SE stage of the $\log_d(N,0,m)$ network does not have
fanout capability, then when $f \leq d^{n-2}$ the network is
$f$-cast strictly non-blocking if
\[
m \geq 
d^{\left\lfloor \frac{n+r}{2} \right\rfloor} +
f\left(d^{\left\lceil \frac{n-r-2}{2} \right\rceil} - 1 \right).
\]
When $f>d^{n-2}$ the network is $f$-cast strictly nonblocking if $m \geq d^{n-1}$.
\label{cor:wang-icc2008}
\ecor
\bp
Recall Remark \ref{rmk:snb-vs-wsnb-remark}:
routing using the window algorithm with window size $t=n$ is 
the same as routing arbitrarily in the network when
the $1\times m$-SE stage cannot fanout. Thus any sufficient
condition for the window algorithm to work is a strictly nonblocking
condition. 
Note that when $t=n$ the dual constraints (DC-1) do not exist!
We construct a feasible solution to the dual linear program as follows.

When $f > d^{n-2}$, set $\gamma_{\bf u} = 1$ for all 
${\bf u} \in \bigcup_{i=1}^{n-1} A_i$,
and all other variables to be $0$. The dual objective value in this 
case is 
\[ \sum_{{\bf u} \in \bigcup_{i=1}^{n-1} A_i} \gamma_{\bf u}
 = \sum_{i=1}^{n-1}|A_i| = \sum_{i=1}^{n-1}(d^{n-i}-d^{n-i-1}) = d^{n-1}-1, 
\] 
and hence one more plane (i.e. $m \geq d^{n-1}$) is sufficient.
Note that this solution is dual feasible, because for $\bf u \in A_0$
there is no $\bf v$ for which $i(\mv u)+j(\mv v) \geq n$. In other words,
there is no dual constraint for which $\mv u \in A_0$.

Next, suppose $f \leq d^{n-2}$. 
Define $q = \left\lfloor \frac{n+r}{2} \right\rfloor+1$.
Note that $r+1\leq q \leq n-1$ in this case;
in particular, 
$kd^{n-q}<d^{r+1}d^{n-q}\leq d^n,$
which implies
$\min\{d^n-k, k(d^{n-q}-1)\} = kd^{n-q}.$
Set 
$\gamma_{\bf u} = 1$ for all ${\bf u}$ with $i(\mv u) \geq n-q+1$
%\begin{equation}
% \gamma_{\bf u} = 
% \begin{cases}
%  1 & \text{ if } i(\mv u) \geq n-q+1\\
%  0 & \text{ otherwise. }
% \end{cases}
%\end{equation}
and 
$\delta_{\bf v} = 1$ for all ${\bf v} \in \bigcup_{j=q}^{n-1} B_j$.
%\begin{equation}
% \delta_{\mv v} = 
% \begin{cases}
%  1 & \text{ if } \mv v \in \bigcup_{j=q}^{n-1} B_j\\
%  0 & \text{ otherwise. }
% \end{cases}
%\end{equation}
All other dual variables are $0$. The solution
is dual feasible because, for any pair $(\mv u, \mv v)$ for which
$i(\mv u)+j(\mv v)\geq n$, we must either have $i(\mv u)\geq n-q+1$
or $j(\mv v) \geq q$ (which is the same as saying 
$\mv v \in \bigcup_{j=q}^{n-1} B_j$).
Recalling Proposition \ref{prop:q-bound}, the dual objective value is
\begin{eqnarray*}
\sum_{\mv u \ : \ i(\mv u)\geq n-q+1} \gamma_{\mv u}
+ \sum_{\mv v \in \bigcup_{j=q}^{n-1} B_j} \delta_{\mv v}
&=& \sum_{i=n-q+1}^{n-1}|A_i|+\left|\bigcup_{j=q}^{n-1} B_j\right|\\
&\leq& d^{q-1}-1+\min\left\{d^n-k,k(d^{n-q}-1) \right\}\\
&=&d^{q-1}-1+k(d^{n-q}-1)\\
&\leq&d^{q-1}+f(d^{n-q}-1)-1,
\end{eqnarray*}
This is an upper bound on the number of blocking planes.
Hence, one more plane is sufficient to route the new (arbitrary) request.
\ep

Because unicast is $1$-cast, 
by setting $r=0$ in the previous corollary we obtain
the following corollary, whose proof was about 5 pages long
in \cite{Hwang98}.
Recall remark \ref{rmk:snb-vs-wsnb-remark} which ensures that our
result is a strictly nonblocking condition rather than a wide-sense
nonblocking one.

\bcor[Theorem 1 in \cite{Hwang98}]
For $\log_d(N,0,m)$ to be unicast strictly nonblocking, it is sufficient
that $m\geq d^{\lceil n/2 \rceil-1} + d^{\lfloor n/2 \rfloor}-1$.
\ecor

Corollary \ref{cor:wang-icc2008} solves the $t=n$ case. 
We will consider $0\leq t<n$ henceforth.
We next specify a family of dual-feasible solutions to the dual-LP
\eqref{eqn:dual-LP}. The main remaining task will be simple calculus
as we pick the best dual-feasible solution depending on 
the parameters $f,n,d,t$ of the problem.

The family of dual-feasible solution is specified with two integral parameters
where $0\leq p \leq n-t-1$ and $n-t\leq q \leq n$.
The parameter $p$ is used to set the variables $\epsilon_{\mv u}$,
$\alpha_w$ and $\beta_{\mv u,w}$, and the parameter $q$ is used to
set the variables $\gamma_{\mv u}$ and $\delta_{\mv v}$.
As we set the variables, we will also verify the feasibility of
the constraints (DC-1) and (DC-2),
and the contributions of those variables to the final objective value.

\bi
\item {\bf Specifying the $\epsilon_{\mv u}$ variables.}
\iffalse
The $\epsilon_{\mv u}$ are defined as follows.
\begin{equation}
 \epsilon_{\mv u} =
  \begin{cases}
    1 & i(\mv u) \geq n-p\\
    0 & \textnormal{otherwise}
  \end{cases}
\label{eqn:epsilon}
\end{equation}
\fi
Set $\epsilon_{\mv u}=1$ if $i(\mv u)\geq n-p$ and $0$ otherwise.
The contribution of the $\epsilon_{\mv u}$ to the objective is
\[
 \sum_{\mv u}f \epsilon_{\mv u} = 
  \sum_{i=n-p}^{n-1} f\sum_{\mv u: i(\mv u)=i} 1 = 
  \sum_{i=n-p}^{n-1} f|A_i|
  = f(d^p-1).                              
\]

\item
{\bf Specifying the $\alpha_w$ and $\beta_{\mv u,w}$ variables.}
Next, we define the $\alpha_w$ and $\beta_{\mv u,w}$.
The constraints (DC-1) with $i(\mv u) \geq n-p$ are already satisfied
by the $\epsilon_{\mv u}$, hence we only need to set the 
$\alpha_w$ and $\beta_{\mv u, w}$ to satisfy (DC-1) when 
$j(w)\geq p+1$.
(If $j(w)\leq p$, then for the constraint to exist we must have
$i(\mv u) \geq n-j(w) \geq n-p$.)
The variables $\alpha_w$ and $\beta_{\mv u,w}$
 are set differently based on three cases as follows.

{\bf Case 1.} If $t \geq \left\lfloor \frac n 2 \right\rfloor $, then set 
$\beta_{{\bf u}, w} = 1$ whenever $p+1 \leq j(w) \leq n-t-1$
and $n-j(w) \leq i(\mv u) \leq n-p-1$,
and set all other $\alpha_w$ and $\beta_{\mv u,w}$ to be $0$.
%\begin{equation}
%\beta_{{\bf u}, w} =
% \begin{cases}
%    1& p+1 \leq j(w) \leq n-t-1, \text{ and }\\
%     & n-j(w) \leq i(\mv u) \leq n-p-1,\\
%    0& \text{otherwise}
% \end{cases}
%\end{equation}
%and set all the $\alpha_w$ to $0$.
It can be verified straightforwardly that all constraints (DC-1)
are satisfied.
Recall that the number of windows $W_w$ for which $j(w)=j$ is
precisely $d^{n-j-t}-d^{n-j-t-1}$.
Thus, the contributions of the $\alpha_w$ and $\beta_{\mv u, w}$
to the dual objective value is
\begin{eqnarray*}
 \sum_{\substack{\mv u, w\\ 
                     p+1 \leq j(w) \leq n-t-1\\ 
                     n-j(w)\leq i(\mv u)\leq n-p-1}} 
     \beta_{\mv u,w}
 &=& \sum_{j=p+1}^{n-t-1}|\{w : j(w) = j\}| 
     \sum_{i=n-j}^{n-p-1}|\{\mv u \in \mathcal A : i(\mv u) = i\}|\\
 &=& \sum_{j=p+1}^{n-t-1} (d^{n-j-t}-d^{n-j-t-1})\sum_{i=n-j}^{n-p-1}|A_i|\\
 &=& \sum_{j=p+1}^{n-t-1} (d^{n-j-t}-d^{n-j-t-1})(d^j-d^p)\\
 &=& (n-t-1-p)(d^{n-t}-d^{n-t-1}) - d^{n-t-1}+d^p.
\end{eqnarray*}
%\[ \sum_{\substack{\mv u, w\\ p< j(w) < n-t}} \beta_{\mv u,w}
% = (n-t-1-p)(d^{n-t}-d^{n-t-1}) - d^{n-t-1}+d^p.
%\]

{\bf Case 2.} When 
$p+1 \leq t \leq \left\lfloor \frac n 2 \right\rfloor-1$, set 
$\beta_{\mv u,w}=1$ whenver $p+1 \leq j(w) \leq t$ and
$n-j(w)\leq i \leq n-p-1$,
%\[
%\beta_{\mv u,w} =
%  \begin{cases} 1& p< j(w) \leq t \text{ and } n-j(w)\leq i < n-p\\
%                0&\text{otherwise}
%  \end{cases}
%\]
and
%\[
% \alpha_w = \begin{cases} 1& t < j < n-t\\0&\text{otherwise}\end{cases}
%\]
$\alpha_w=1$ for $t+1 \leq j(w) \leq n-t-1$, and all other
$\alpha_w$ and $\beta_{\mv u,w}$ to be $0$.
All constraints (DC-1) are thus satisfied.
The $\alpha_w$'s and $\beta_{\mv u,w}$'s contributions to the objective is
\begin{eqnarray*}
 \sum_{\substack{w\\t < j(w) < n-t}} d^t\alpha_w
 + 
 \sum_{\substack{\mv u, w\\ 
                 p+1 \leq j(w) \leq t\\ 
                 n-j(w)\leq i(\mv u)\leq n-p-1}} \beta_{\mv u, w}
 &=&
 \sum_{j=t+1}^{n-t-1} d^t \cdot |\{ w : j(w) = j \}| +\\
 &&
 \sum_{j=p+1}^{t}|\{w : j(w) = j\}| 
 \sum_{i=n-j}^{n-p-1}|\{\mv u \in \mathcal A : i(\mv u) = i\}|\\
 &=& 
 \sum_{j=t+1}^{n-t-1}d^t(d^{n-j-t}-d^{n-j-t-1}) + \\
 && \sum_{j=p+1}^t (d^j-d^p)(d^{n-j-t}-d^{n-j-t-1})\\
 &=& (t-p)(d^{n-t}-d^{n-t-1}) + d^{n+p-2t-1}-d^t.
\end{eqnarray*}

{\bf Case 3.} When $t \leq p$ (which is $\leq n-t-1$), set
$\alpha_w=1$ for $p+1 \leq j(w) \leq n-t-1$ 
%\begin{equation}
%\alpha_w =
% \begin{cases}
%     1 & p+1 \leq j(w) \leq n-t-1\\
%     0 & \text{otherwise}
% \end{cases}
%\end{equation}
and all the $\beta_{\mv u,w}$ to be zero. Again, the feasibility of
the constraints (DC-1) is easy to verify. The contribution
to the objective value is 
\begin{eqnarray*}
 \sum_{\substack{w\\p < j(w) < n-t}} d^t\alpha_w
&=& \sum_{j=p+1}^{n-t-1} d^t \cdot |\{w : j(w) = j\}|\\
&=& \sum_{j=p+1}^{n-t-1} d^t(d^{n-j-t}-d^{n-j-t-1}) \\
&=& d^{n-p-1}-d^t.
\end{eqnarray*}

\item
{\bf Specifying the $\gamma_{\mv u}$ and $\delta_{\mv v}$ variables.}
Here, there are two cases

When $q = n-t$, set $\delta_{\mv v}=1$ for all 
$\mv v \in \bigcup_{j=n-t}^{n-1} B_j$
and all $\gamma_{\mv u}=0$. The dual-objective contribution in this case
is 
\[ \sum_{\mv v \in \bigcup_{j=n-t}^{n-1} B_j} \delta_{\mv v} = 
   \left|\bigcup_{j=n-t}^{n-1} B_j\right| = d^t - k.
\]
When $n-t+1\leq q \leq n$, define
$\delta_{\mv v}=1$ for all $\mv v \in \bigcup_{j=q}^{n-1} B_j$,
$\gamma_{\mv u}=1$ for all $\mv u$ such that 
$n-q+1\leq i(\mv u) \leq n-p-1$,
and all other $\delta_{\mv v}$ and $\gamma_{\mv u}$ are set to be zero.
\iffalse
\begin{equation}
\delta_{\mv v} = 
 \begin{cases}
  1 & \mv v \in \bigcup_{j=q}^{n-1} B_j\\
  0 & \text{otherwise}
 \end{cases}
\end{equation}
and
\begin{equation}
\gamma_{\mv u} = 
 \begin{cases}
  1 & n-q+1\leq i(\mv u) \leq n-p-1\\
  0 & \text{otherwise}
 \end{cases}
\end{equation}
\fi
From Proposition \ref{prop:q-bound}, 
the total contribution of the $\gamma_{\mv u}$ and 
$\delta_{\mv v}$ to the dual-objective is at most
\begin{eqnarray*}
 \sum_{\substack{\mv u\\n-q+1\leq i(\mv u) \leq n-p-1}} \gamma_{\mv u} +
 \sum_{\mv v \in \bigcup_{j=q}^{n-1} B_j} \delta_{\mv v} 
 &=& \sum_{i=n-q+1}^{n-p-1}|\{\mv u : i(\mv u)=i\}| 
     + \left|\bigcup_{j=q}^{n-1} B_j\right| \\
 &\leq& \sum_{i=n-q+1}^{n-p-1}|A_i| + \min\{d^t-k, k(d^{n-q}-1)\} \\
 &=& d^{q-1} - d^p + \min\{d^t-k, k(d^{n-q}-1)\}.
\end{eqnarray*}
The feasibility of all the constraints (DC-2) is easy to verify.
\ei

Define the ``cost'' $c(k,p,q)$ to be
the total contribution of all variables to the dual-objective value.
We summarize the values of $c(k,p,q)$ in 
Figure~\ref{fig:cost}.
We just proved the following.

\begin{figure*}[t]
\footnotesize
\bbox \centerline{\bf The objective value $c(k,p,q)$} \ebox
For $t \geq \left\lfloor \frac n 2 \right\rfloor$ and $q=n-t$,
\[
c(k,p,q) = 
   f(d^p-1)+(n-t-1-p)(d^{n-t}-d^{n-t-1}) - d^{n-t-1}+d^p + d^t-k.
\]
For $t \geq \left\lfloor \frac n 2 \right\rfloor$ and $q>n-t$,
\[
c(k,p,q) = 
   f(d^p-1)+(n-t-1-p)(d^{n-t}-d^{n-t-1}) - d^{n-t-1}+
   d^{q-1} + \min\{d^t-k, k(d^{n-q}-1)\}.
\]
For $p+1 \leq t \leq \left\lfloor \frac n 2 \right\rfloor-1$ and $q=n-t$
\[
c(k,p,q) = 
   f(d^p-1)+(t-p)(d^{n-t}-d^{n-t-1}) + d^{n+p-2t-1}-k.
\]
For $p+1 \leq t \leq \left\lfloor \frac n 2 \right\rfloor-1$ and $q>n-t$
\[
c(k,p,q) = 
   f(d^p-1)+(t-p)(d^{n-t}-d^{n-t-1}) + d^{n+p-2t-1}-d^t + d^{q-1}-d^p+
   \min\{d^t-k, k(d^{n-q}-1)\}.
\]
For $t \leq p$ and $q=n-t$,
\[ c(k,p,q) = f(d^p-1)+d^{n-p-1}-k. \]
For $t \leq p$ and $q>n-t$,
\[
c(k,p,q) = 
   f(d^p-1)+d^{n-p-1}-d^t+d^{q-1}-d^p+ \min\{d^t-k, k(d^{n-q}-1)\}.
\]
\caption{The dual objective value of the family of dual-feasible solutions.}
\label{fig:cost}
\end{figure*}

\bthm
The above family of solutions is feasible for the dual linear program 
\eqref{eqn:dual-LP}
with objective value equal to $c(k,p,q)$.
Consequently, for the network $\log_d(N,0,m)$ to be wide-sense nonblocking under the
window algorithm with window size $d^t$, it is sufficient that
\beq
  m \geq 1+\max_{1\leq k\leq \min(f,d^t)} \min_{p,q} c(k,p,q). 
\label{eqn:sufcond}
\eeq
\label{thm:mainlog}
\ethm

\subsection{Selecting the best dual-feasible solution}

It is a very straightforward though somewhat analytically tedious task 
to derive the best possible sufficient condition using 
Theorem \ref{thm:mainlog}.
The idea is, for a given $k \leq \min(f,d^t)$, we first choose
$p=p_k,q=q_k$ so that $c(k,p_k,q_k)$ is as small as possible.
Then, derive an upperbound $C(t,f) \geq \max_k c(k,p_k,q_k)$.
The sufficient condition is then $m \geq C(t,f)+1$.

\begin{figure*}[t]
\footnotesize
\bbox \centerline{\bf The upper-bound $C(t,f)$} \ebox
To shorten the notations, let $r = \lfloor \log_d f \rfloor$.
\[
C(t,f) = 
 \begin{cases}
f\left(d^{\left\lceil\frac{n-r}{2}\right\rceil-1}-1\right) 
 + d^{n-\left\lceil\frac{n-r}{2}\right\rceil}-1
   & t < \left\lfloor \frac n 2 \right\rfloor, r \leq n-2t-1  \\
t(d-1)d^{n-t-1} + d^{n-2t-1}-1
   & t < \left\lfloor \frac n 2 \right\rfloor, r \geq n-2t \\
[(n-t-1)(d-1)-1]d^{n-t-1}
+ d^t-(d-1)d^{2t-n-1}
   & t \geq \left\lfloor \frac n 2 \right\rfloor, r \geq n-t \\
 f\left(d^{n-t-r-1} - 1\right) + [r(d-1)-1]d^{n-t-1}
+d^{n-t-r-1}+d^t-(d-1)d^{2t-n-1}
   & t \geq \left\lfloor \frac n 2 \right\rfloor, \text{ and } \\
   & 2t-n-2 < r \leq n-t-1\\
 f\left(d^{n-t-r-1} - 1\right) + [r(d-1)-1]d^{n-t-1}
 + d^{\left\lfloor \frac{n+r}{2} \right\rfloor} 
+f\left(d^{n-\left\lfloor \frac{n+r}{2} \right\rfloor -1}-1\right)
   & t \geq \left\lfloor \frac n 2 \right\rfloor, \text{ and } \\
   & r \leq \min(2t-n-2,n-t-1)
 \end{cases}
\]
\caption{We show in Theorem \ref{thm:general}
         that $C(t,f) \geq \max_k \min_{p,q} c(k,p,q)$}
\label{fig:C(t,f)}
\end{figure*}

We first need a technical lemma.

\begin{lmm}
Let $d,n,k$ be positive integers, and $x=\lfloor \log_d k \rfloor$.
Then, the following function
\begin{equation}
 h(k) = 
d^{\left\lfloor\frac{n+x}{2}\right\rfloor} +
k\left(d^{n-\left\lfloor\frac{n+x}{2}\right\rfloor-1}-1\right)
\label{eqn:h(k)}
\end{equation}
is non-decreasing in $k$.
\label{lmm:technical1}
\end{lmm}
\bp
We induct on $k$. The inequality trivially holds when $k=1$.
Consider $k>2$. First, suppose $k$ is not an exact power of $d$, i.e.
$k>d^x$. In this case, we have
\begin{eqnarray*}
h(k-1) &=&d^{\left\lfloor\frac{n+x}{2}\right\rfloor} +
(k-1)\left(d^{n-\left\lfloor\frac{n+x}{2}\right\rfloor-1}-1\right)\\
&\leq& d^{\left\lfloor\frac{n+x}{2}\right\rfloor} +
k\left(d^{n-\left\lfloor\frac{n+x}{2}\right\rfloor-1}-1\right) 
= h(k).
\end{eqnarray*}
Second, consider the case when $k=d^x$. 
It can be verified that,
no matter what the parities of $n$ and $x$ are, the multiset
$\left\{\left\lfloor\frac{n+x-1}{2}\right\rfloor, 
\left\lceil\frac{n+x-1}{2}\right\rceil \right\}$ 
is exactly equal to the multiset
$\left\{\left\lfloor\frac{n+x}{2}\right\rfloor, 
\left\lceil\frac{n+x}{2}\right\rceil-1\right\}$.
Thus, noting that $\lfloor \log_d(k-1) \rfloor = x-1$, we have
\begin{eqnarray*}
h(k-1)&=&
d^{\left\lfloor\frac{n+x-1}{2}\right\rfloor} +
(k-1)\left(d^{n-\left\lfloor\frac{n+x-1}{2}\right\rfloor-1}-1\right)\\
&=&
d^{\left\lfloor\frac{n+x-1}{2}\right\rfloor} +
(d^x-1)\left(d^{\left\lceil \frac{n-x-1}{2}\right\rceil}-1\right)\\
&=&
d^{\left\lfloor\frac{n+x-1}{2}\right\rfloor} +
d^{\left\lceil\frac{n+x-1}{2}\right\rceil} -
d^{\left\lceil\frac{n-x-1}{2}\right\rceil} - d^x+1\\
&=&
d^{\left\lfloor\frac{n+x}{2}\right\rfloor} +
d^{\left\lceil\frac{n+x}{2}\right\rceil-1} -
d^{\left\lceil\frac{n-x-1}{2}\right\rceil} - d^x+1\\
&\leq&
d^{\left\lfloor\frac{n+x}{2}\right\rfloor} +
d^{\left\lceil\frac{n+x}{2}\right\rceil-1} - d^x\\
&=&
d^{\left\lfloor\frac{n+x}{2}\right\rfloor} +
d^x\left(d^{n-\left\lfloor\frac{n+x}{2}\right\rfloor-1}-1\right)\\
&=&h(k).
\end{eqnarray*}
\ep

\bthm
The $\log_d(N,0,m)$ network is nonblocking under the window algorithm with
window size $d^t$ if $m\geq 1+C(t,f)$ where $C(t,f)$ is defined
in Figure \ref{fig:C(t,f)}.
\label{thm:general}
\ethm
\bp
Consider $5$ cases in the definition of $C(t,f)$. 
We specify for each $k$ how to set the values 
$p_k$ and $q_k$.
%And then, we verify that $c(k,p_k,q_k) \leq C(t,f)$.
The straightforward task of verifying that $c(k,p_k,q_k) \leq C(t,f)$
is mostly omitted due to space constraint, except for situations when
it is tricky to verify.

\bi
\item {\bf Case 1:} $t < \left\lfloor \frac n 2 \right\rfloor, r \leq n-2t-1$.
For any $k$, choose $p_k = \left\lceil \frac{n-r}{2}-1\right\rceil$
and $q_k = n-t$. 
\iffalse
Noting that $p_k\geq t$ and $k\geq 1$, we have
\[ c(k,p_k,q_k) = 
  f\left(d^{\left\lceil\frac{n-r}{2}\right\rceil-1}-1\right) 
 + d^{n-\left\lceil\frac{n-r}{2}\right\rceil}-k \leq C(t,f).
\]
\fi

\item
{\bf Case 2:} $t < \left\lfloor \frac n 2 \right\rfloor, r \geq n-2t$.
For any $k$, set $p_k=0$ and $q_k=n-t$. 
\iffalse
Then, 
\begin{eqnarray*} 
c(k,p_k,q_k) &=& t(d^{n-t}-d^{n-t-1})+d^{n-2t-1}-k \\
&  \leq &t(d^{n-t}-d^{n-t-1})+d^{n-2t-1}-1 \\
&=& C(t,f).
\end{eqnarray*}
\fi

\item
{\bf Case 3:} $t \geq \left\lfloor \frac n 2 \right\rfloor, r \geq n-t$.
This case is a little trickier analytically. 
Define $x = \lfloor \log_d k\rfloor$.
We set $p_k$ and $q_k$ differently depending on how large $x$ is,
so that the inequality $c(k,p_k,q_k) \leq C(t,f)$ always holds.

If $0\leq x \leq 2t-n-2$, which can only hold when $t\geq \frac{n+1}{2}$,
then set $q_k = \left\lfloor\frac{n+x}{2}\right\rfloor+1$ and $p_k=0$.
%\iffalse
Note that $q_k>n-t$ and $x+1+n-q_k<t$. Thus $kd^{n-q_k}<d^t$. 
Recall from Lemma \ref{lmm:technical1} that function $h(k)$ 
defined in \eqref{eqn:h(k)} is non-increasing,
and the fact that in this case $k \leq d^{x+1}-1 \leq d^{2t-n-1}-1$,
we have
\begin{eqnarray*}
c(k,p_k,q_k) 
&=& [(n-t-1)(d-1)-1]d^{n-t-1} + d^{q_k-1} + 
\min\{d^t-k,k(d^{n-q_k}-1)\}\\
&=& [(n-t-1)(d-1)-1]d^{n-t-1} + d^{q_k-1} + k(d^{n-q_k}-1)\\
&=& [(n-t-1)(d-1)-1]d^{n-t-1} + h(k)\\
&\leq& [(n-t-1)(d-1)-1]d^{n-t-1} + h(d^{2t-n-1}-1)\\
&=& [(n-t-1)(d-1)-1]d^{n-t-1} +d^{t-1} + 
(d^{2t-n-1}-1)(d^{n-t}-1)\\
&<& [(n-t-1)(d-1)-1]d^{n-t-1} +d^{t-1} + 
d^{2t-n-1}(d-1)(d^{n-t}-1)\\
&=& [(n-t-1)(d-1)-1]d^{n-t-1} +d^t-(d-1)d^{2t-n-1}\\
&=& C(t,f).
\end{eqnarray*}
%We bound the second part of $c(k,p_k,q_k)$:
%\begin{eqnarray*}
%d^{q_k-1} + \min\{d^t-k,k(d^{n-q_k}-1)\}
%&=&d^{q_k-1}+k(d^{n-q_k}-1)\\
%&\leq&d^{\left\lfloor\frac{n+x}{2}\right\rfloor}+
%      (d^{x+1}-1)\left(d^{n-\left\lfloor\frac{n+x}{2}\right\rfloor-1}-1\right)
%\end{eqnarray*}
%Let $h(x)$ denote the last expression above.
%Then, $h(x)$ is increasing as a function of the non-negative integer
%$x$ whenever $x\leq n-1$.
%Hence, 
%\begin{eqnarray*}
% h(x) &\leq& h(2t-n-2) \\
%&=& d^{t-1} + (d^{2t-n-1}-1)(d^{n-t}-1)\\
%&<& d^{t-1} + d^{2t-n-1}(d-1)(d^{n-t}-1)\\
%&=& d^t-(d-1)d^{2t-n-1}.
%\end{eqnarray*}
%Consequently, $c(k,p_k,q_k) \leq C(t,f)$ as desired.
%\fi

If $x=2t-n-1$ and $k\leq d^{x+1}-d^x$, then set
$q_k = \left\lfloor\frac{n+x}{2}\right\rfloor+1=t$ and $p_k=0$.
\iffalse
Note that $q_k=t>n-t$ because $x\geq 0$.
In this case, 
\begin{eqnarray*}
&& c(k,p_k,q_k) \\
%&=& [(n-t-1)(d-1)-1]d^{n-t-1} + d^{q_k-1} + \min\{d^t-k,k(d^{n-q_k}-1)\}\\
%&=& [(n-t-1)(d-1)-1]d^{n-t-1} + d^{q_k-1} + k(d^{n-q_k}-1)\\
&=& [(n-t-1)(d-1)-1]d^{n-t-1} + h(k)\\
&\leq& [(n-t-1)(d-1)-1]d^{n-t-1} + \\
&&h(d^{2t-n}-d^{2t-n-1})\\
&=&[(n-t-1)(d-1)-1]d^{n-t-1} + d^{t-1} + \\
&&(d^{2t-n}-d^{2t-n-1})(d^{n-t}-1)\\
&=&[(n-t-1)(d-1)-1]d^{n-t-1} + \\
&&d^t-(d-1)d^{2t-n-1}\\
&=&C(t,f).
\end{eqnarray*}
\fi
If $x=2t-n-1$ and $k\geq d^{x+1}-d^x+1$, then set
set $q_k = n-t$ and $p_k=0$.
\iffalse
Then, 
\begin{eqnarray*}
&&c(k,p_k,q_k) \\
&=& [(n-t-1)(d-1)-1]d^{n-t-1} + 1 + d^t-k\\
&\leq& [(n-t-1)(d-1)-1]d^{n-t-1} + 1 + \\
&&d^t - d^{x+1}+d^x-1\\
&=& [(n-t-1)(d-1)-1]d^{n-t-1} + \\
&&d^t - (d-1)d^{2t-n-1}\\
&=&C(t,f).
\end{eqnarray*}
\fi
%
Finally, when $x\geq 2t-n$, we again set $q_k=n-t$ and $p_k=0$.
\iffalse
Note that $k\geq d^x\geq d^{2t-n}>(d-1)d^{2t-n-1}$. Thus,
$d^t-k<d^t-(d-1)d^{2t-n-1}-1$ and $c(k,p_k,q_k) \leq C(t,f)$ follows
straightforwardly.
\fi

\item
{\bf Case 4:} $t \geq \left\lfloor \frac n 2 \right\rfloor, 
2t-n-2 < r \leq n-t-1$.
Note that this case can only happen when $t\leq 2n/3$.
In particular, if $t>2n/3$ and $r\leq n-t-1$ we would be in case 5.
Set $p_k=n-t-r-1$ and $q_k = \left\lfloor \frac{n+x}{2} \right\rfloor+1$.
%\iffalse
Proving $c(k,p_k,q_k) \leq C(t,f)$ is almost identical to Case 3 where
we consider different ranges of $x =\left\lfloor \log_d k \right\rfloor$.
%\fi

\item
{\bf Case 5:} $t \geq \left\lfloor \frac n 2 \right\rfloor, 
r \leq \min(2t-n-2,n-t-1)$.
Set $p_k=n-t-r-1$ and $q_k = \left\lfloor \frac{n+x}{2} \right\rfloor+1$.
%\iffalse
Showing $c(k,p_k,q_k) \leq C(t,f)$ is similar to Case 3.
The only slight variation is, instead of bounding $k\leq d^{x+1}-1$
we apply $k\leq f$ directly. The function $h(k)$ is then bounded by
$h(f)$.
Furthermore, we do not have to consider the cases when $x\geq 2t-n-1$
because $x\leq r \leq 2t-n-2$.
\ei
%\fi
\ep

\subsection{Some quick consequences of Theorem \ref{thm:general}}

All we have to do is to plug in the parameters $t$ and $f$
and compute $1+C(t,f)$ to get the following results.

%\bcor[A generalized version of Theorem 1 in \cite{Lea90}]
%For $\log_d(N,0,m)$ to be unicast strictly nonblocking, it is sufficient that
%$m \geq d^{\lfloor n/2 \rfloor}+d^{\lceil n/2\rceil-1}-1$.
%\ecor
%\bp $C(t,1) = d^{\lfloor n/2 \rfloor}+d^{\lceil n/2\rceil-1}-1$.  \ep

\begin{cor}[Theorem 4 in \cite{Wang07}]
Let $r = \lfloor \log_d f \rfloor$.
The network $\log_d(N,0,m)$ is $f$-cast strictly non-blocking if
\[ m \geq f\left( d^{\left\lceil \frac{n-r}{2} \right\rceil-1} - 1\right) 
   + d^{n - \left\lceil \frac{n-r}{2} \right\rceil}. 
\]
\label{cor:wang07}
\end{cor}
\bp
This corresponds to the $t=0$ case of the window algorithm, which becomes
a strictly nonblocking condition as noted earlier.
\[ C(0,f) =
f\left( d^{\left\lceil \frac{n-r}{2} \right\rceil-1} - 1\right) 
   + d^{n - \left\lceil \frac{n-r}{2} \right\rceil}-1.
\]
\ep

The following result took about $6$ pages in 
\cite{DBLP:journals/tcom/Danilewicz07} to be proved
(in two theorems) with combinatorial reasoning.
The result is on the general multicast case, without the fanout restriction
$f$. In our setting, we can simply set $f=N=d^n$.
In fact, even though the corollary states exactly the same results
as in \cite{DBLP:journals/tcom/Danilewicz07}, the statement is simpler.
\bcor[Theorems $1$ and $2$ in \cite{DBLP:journals/tcom/Danilewicz07}]
The $d$-ary multi-log network $\log_d(N,0,m)$
is wide-sense nonblocking with respect to the window algorithm
with window size $d^t$ if 
\[ 
m \geq
\begin{cases}
  d^{n-2t-1}+td^{n-t-1}(d-1) & \text{ when } t\leq \left\lfloor \frac n 2 \right\rfloor -1,\\
  d^{n-t-1}[(d-1)(n-t-1)-1]&\\
  +d^t-d^{2t-n-1}(d-1)+1 &
  \text{ when } t \geq \left\lfloor \frac n 2 \right\rfloor.
\end{cases}
\]
\label{thm:1and2}
\ecor
%\bp
%Note that $r=n$. We have
%\[ C(t,d^n) = d^{n-2t-1}+td^{n-t-1}(d-1)-1, \text{ when }
%t\leq \left\lfloor \frac n 2 \right\rfloor -1, 
%\] 
%and
%\[ C(t,d^n) = d^{n-t-1}[(d-1)(n-t-1)-1]+d^t-d^{2t-n-1}(d-1), 
%\]
%otherwise.
%\ep
%\input{cf-fcast-WSNB}

%%%%%%%%%%%%%%%%%%%%%%%%%%%%%%%%%%%%%%%%%%%%%%%%%%%%%%%%%%%%%%%%%%%%%%%%
\section{Analyzing crosstalk-free $f$-cast wide-sense nonblocking multilog networks}
\label{sec:cf-fcast-WSNB}

When the multi-log architecture is employed to design a photonic switch,
each $2\times 2$ switching element (SE) needs to be replaced by a functionally
equivalent optical component.
For instance, when $d=2$ we can use so-called {\em directional couplers}
as SEs \cite{stern-book, DBLP:conf/globecom/WuT06, mukherjee-book}.
However, directional couplers and many other optical switching
elements suffer from optical crosstalk between interfering channels, which
is one of the major obstacles in designing cost-effective switches
\cite{vaez-lea-1998, Chinni, Li}.
To cope with crosstalk, the {\em crosstalk-free constraint} is a common
requirement, which states that no two routes can share a common SE
\cite{Lea90,vaez-lea-2000,maier-pattavina-2001,966007, Chinni, Li,wang-icc2008}.

Thanks to Proposition \ref{prop:unicast-node-block}, to analyze crosstalk-free
$f$-cast wide-sense nonblocking multilog networks under the window algorithm, basically
all we have to do is to replace the constraint $i+j\geq n$ by the constraint
$i+j\geq n-1$. That was essentially the only difference between 
two Propositions \ref{prop:unicast-node-block} and
\ref{prop:unicast-link-block}. Replacing $n$ by $n-1$ leads to 
changes in the final formula for the required number of Banyan planes.
Deriving the formulas is relatively straightforward but also takes
takes some (straightforward) calculus effort and thus we do so here. 
The overall outline of the analysis, however, is identical and we can 
reuse much of the analysis for the non crosstalk-free case.

\subsection{Setting up the linear program and its dual}

We use identical notations as in the previous section.
The following lemma is the crosstalk-free analog of Lemma
\ref{lmm:LP}.

\blmm
For each input $\mv u \in \mathcal A$ and
each $w\in [d^{n-t}-1]$ such that $i(\mv u)+j(w)\geq n-1$,
define a variable $x_{\mv u,w}$.
Also,
for each input $\mv u \in \mathcal A$ and
each output $\mathbf v\in W_0-B$ such that
$i(\mv u)+j(\mv v) \geq n-1$,
define a variable $x_{\mv u,\mv v}$.
Then, the number of Banyan planes blocking $(\mv a, B)$ is upperbounded
by the optimal value of the linear program \eqref{eqn:LP}, whose dual
is \eqref{eqn:dual-LP}.
\elmm

We next derive some quick consequences of the formulation.
\bcor[Theorem III.1 in \cite{wang-icc2008}]
Let $r = \lfloor \log_d f \rfloor$.
Suppose the $1\times m$-SE stage of the $\log_d(N,0,m)$ network does not 
have fanout capability, then when $f \leq d^{n-2}(d-1)$ the network is
crosstalk-free $f$-cast strictly non-blocking if
\[
m \geq 
d^{\left\lfloor \frac{n+r+1}{2} \right\rfloor} +
f\left(d^{\left\lceil \frac{n-r-1}{2} \right\rceil} - 1 \right).
\]
When $f>d^{n-2}(d-1)$ the network is $f$-cast strictly nonblocking if 
$m \geq d^n-d^{n-2}(d-1)$.
\label{cor:wang-icc2008-2}
\ecor
\bp
Routing using the window algorithm with window size $t=n$ is 
the same as routing arbitrarily in the network when
the $1\times m$-SE stage cannot fanout. Thus any sufficient
condition for the window algorithm to work is an strictly nonblocking
condition. 
Note that when $t=n$ the dual constraints (DC-1) do not exist.
Consider a solution to the dual LP as follows.

When $f > d^{n-2}(d-1)$, consider two cases.
If $k > d^{n-2}(d-1)$, set $\delta_{\mv v} = 1$ for all 
$\mv v \in \bigcup_{j=0}^{n-1}B_j$ and all other variables to be $0$.
Then, the dual objective value is
\[ \sum_{\mv v \in \bigcup_{j=0}^{n-1}B_j} \delta_{\mv v}
   = \left|\bigcup_{j=0}^{n-1}B_j\right|=d^n-k \leq d^n-d^{n-2}(d-1)-1. 
\]
Thus, in this case $d^n-d^{n-2}(d-1)$ Banyan planes is sufficient.
Next, suppose $k \leq d^{n-2}(d-1)$, in which case $kd < d^n$.
Set $\gamma_{\mv u}=1$ for all $\mv u$ with $i(\mv u) \geq 1$,
$\delta_{\mv v}=1$ for all $\mv v \in B_{n-1}$, 
and all other variables to be $0$. The solution is dual-feasible
with dual objective value 
\begin{eqnarray*}
\sum_{\mv u: i(\mv u) \geq 1}\gamma_{\mv u} + 
\sum_{\mv v \in B_{n-1}} \delta_{\mv v}
&=&
 \sum_{i=1}^{n-1}|A_i| +|B_{n-1}| \\
&\leq& \sum_{i=1}^{n-1}(d^{n-i}-d^{n-i-1})+\min\{d^n-k,k(d-1)\}\\
&=&d^{n-1}-1+k(d-1)\\
&\leq&d^{n-1}+d^{n-2}(d-1)^2-1\\
&=&d^n-d^{n-2}(d-1)-1.
\end{eqnarray*}
and thus again $d^n-d^{n-2}(d-1)$ Banyan planes is sufficient.

Next, consider the case when $f\leq d^{n-2}(d-1)$. 
In this case $r\leq n-2$.
Let $p = \left\lceil \frac{n-r-1}{2}\right\rceil$.
Then, $1 \leq p \leq \left\lceil \frac{n-1}{2} \right\rceil$.
Furthermore, 
\[ kd^p\leq fd^p<d^{r+1}d^{\left\lceil \frac{n-r-1}{2}\right\rceil}
   = d^{\left\lceil \frac{n+r+1}{2}\right\rceil}
   \leq d^n.
\]
Set $\gamma_{\mv u}=1$ for all $\mv u$ with $i(\mv u) \geq p$,
$\delta_{\mv v}=1$ for all $\mv v \in \bigcup_{j=n-p}^{n-1} B_j$
and all other variables to be $0$. 
The solution
is dual feasible because, for any pair $(\mv u, \mv v)$ for which
$i(\mv u)+j(\mv v)\geq n-1$, we must either have $i(\mv u)\geq p$
or $j(\mv v) \geq n-p$ (which is the same as saying 
$\mv v \in \bigcup_{j=n-p}^{n-1} B_j$).
Recalling Proposition \ref{prop:q-bound} and the fact
that $kd^p<d^n$ shown above,
the dual objective value is
\begin{eqnarray*}
\sum_{\mv u : i(\mv u) \geq p}\gamma_{\mv u} +
\sum_{\mv v \in \bigcup_{j=n-p}^{n-1} B_j} \delta_{\mv v}
&=&
 \sum_{i=p}^{n-1}|A_i| +\left|\bigcup_{j=n-p}^{n-1} B_j\right| \\
&\leq& \sum_{i=p}^{n-1}(d^{n-i}-d^{n-i-1})+\min\{d^n-k,k(d^p-1)\}\\
&=&d^{n-p}-1+k(d^p-1)\\
&\leq&d^{n-p}+f(d^p-1)-1
\end{eqnarray*}
Hence, in this case $d^{n-p}+f(d^p-1)$ is a sufficient number
of Banyan planes.
\ep

\subsection{Specifying a family of dual-feasible solutions}

The family of dual-feasible solution is specified with two integral parameters
where $0\leq p \leq n-t-1$ and $n-t\leq q \leq n$.
The parameter $p$ is used to set the variables $\epsilon_{\mv u}$,
$\alpha_w$ and $\beta_{\mv u,w}$, and the parameter $q$ is used to
set the variables $\gamma_{\mv u}$ and $\delta_{\mv v}$.
As we set the variables, we will also verify the feasibility of
the constraints (DC-1) and (DC-2),
and the contributions of those variables to the final objective value.

\bi
\item
{\bf Specifying the $\epsilon_{\mv u}$ variables.}
\iffalse
The $\epsilon_{\mv u}$ are defined as follows.
\begin{equation}
 \epsilon_{\mv u} =
  \begin{cases}
    1 & i(\mv u) \geq n-p\\
    0 & \textnormal{otherwise}
  \end{cases}
\label{eqn:epsilon1}
\end{equation}
\fi
Set $\epsilon_{\mv u}=1$ if $i(\mv u)\geq n-p$ and $0$ otherwise.
The contribution of the $\epsilon_{\mv u}$ to the objective is
\[
 \sum_{\mv u}f \epsilon_{\mv u} = 
  \sum_{i=n-p}^{n-1} f\sum_{\mv u: i(\mv u)=i} 1 = 
  \sum_{i=n-p}^{n-1} f|A_i|
  = f(d^p-1).                              
\]

\item
{\bf Specifying the $\alpha_w$ and $\beta_{\mv u,w}$ variables.}
Next, we define the $\alpha_w$ and $\beta_{\mv u,w}$.
The constraints (DC-1) with $i(\mv u) \geq n-p$ are already satisfied
by the $\epsilon_{\mv u}$, hence we only need to set the 
$\alpha_w$ and $\beta_{\mv u, w}$ to satisfy (DC-1) when 
$j(w)\geq p$.
(If $j(w)\leq p-1$, then for the constraint to exist we must have
$i(\mv u) \geq n-1-j(w) \geq n-p$.)
The variables $\alpha_w$ and $\beta_{\mv u,w}$
 are set differently based on three cases as follows.

{\bf Case 1.} If $t \geq \left\lceil \frac n 2 \right\rceil $, then set 
$\beta_{{\bf u}, w} = 1$ whenever $p \leq j(w) \leq n-t-1$
and $n-1-j(w) \leq i(\mv u) \leq n-p-1$,
and set all other $\alpha_w$ and $\beta_{\mv u,w}$ to be $0$.
%\begin{equation}
%\beta_{{\bf u}, w} =
% \begin{cases}
%    1& p \leq j(w) \leq n-t-1, \text{ and }\\
%     & n-1-j(w) \leq i(\mv u) \leq n-p-1,\\
%    0& \text{otherwise}
% \end{cases}
%\end{equation}
%and set all the $\alpha_w$ to $0$.
It can be verified straightforwardly that all constraints (DC-1)
are satisfied.
Thus, the contributions of the $\alpha_w$ and $\beta_{\mv u, w}$
to the dual objective value is
\begin{eqnarray*}
 \sum_{\substack{\mv u, w\\ 
                     p \leq j(w) \leq n-t-1\\ 
                     n-1-j(w)\leq i(\mv u)\leq n-p-1}} 
     \beta_{\mv u,w}
% &=& \sum_{j=p}^{n-t-1}|\{w : j(w) = j\}| 
%     \sum_{i=n-1-j}^{n-p-1}|\{\mv u \in \mathcal A : i(\mv u) = i\}|\\
 &=& \sum_{j=p}^{n-t-1} (d^{n-j-t}-d^{n-j-t-1})\sum_{i=n-1-j}^{n-p-1}|A_i|\\
% &=& \sum_{j=p}^{n-t-1} (d^{n-j-t}-d^{n-j-t-1})(d^{j+1}-d^p)\\
 &=& (n-t-p)(d^{n-t+1}-d^{n-t}) - d^{n-t}+d^p.
\end{eqnarray*}
The second equality follows from the fact that the number of windows 
$W_w$ for which $j(w)=j$ is precisely $d^{n-j-t}-d^{n-j-t-1}$.
%\[ \sum_{\substack{\mv u, w\\ p< j(w) < n-t}} \beta_{\mv u,w}
% = (n-t-1-p)(d^{n-t}-d^{n-t-1}) - d^{n-t-1}+d^p.
%\]

{\bf Case 2.} When 
$p+1 \leq t \leq \left\lceil \frac n 2 \right\rceil-1$, set 
$\beta_{\mv u,w}=1$ whenver $p \leq j(w) \leq t-1$ and
$n-1-j(w)\leq i \leq n-p-1$,
%\[
%\beta_{\mv u,w} =
%  \begin{cases} 1& p\leq j(w) \leq t-1 \text{ and } \\ 
%                 & n-1-j(w)\leq i < n-p\\
%                0&\text{otherwise}
%  \end{cases}
%\]
and
%\[
% \alpha_w = \begin{cases} 1& t \leq  j < n-t\\0&\text{otherwise}\end{cases}
%\]
$\alpha_w=1$ for $t \leq j(w) \leq n-t-1$, and all other
$\alpha_w$ and $\beta_{\mv u,w}$ to be $0$.
All constraints (DC-1) are thus satisfied.
The $\alpha_w$'s and $\beta_{\mv u,w}$'s contributions to the objective is
\begin{eqnarray*}
 & & \sum_{\substack{w\\t \leq j(w) < n-t}} d^t\alpha_w
 + 
 \sum_{\substack{\mv u, w\\ 
                 p \leq j(w) \leq t-1\\ 
                 n-1-j(w)\leq i(\mv u)\leq n-p-1}} \beta_{\mv u, w}\\
% &=&
% \sum_{j=t}^{n-t-1} d^t \cdot |\{ w : j(w) = j \}| +\\
%&&
% \sum_{j=p}^{t-1}|\{w : j(w) = j\}| 
% \sum_{i=n-1-j}^{n-p-1}|\{\mv u \in \mathcal A : i(\mv u) = i\}|\\
 &=&
 \sum_{j=t}^{n-t-1}d^t(d^{n-j-t}-d^{n-j-t-1}) + 
 \sum_{j=p}^{t-1} (d^{n-j-t}-d^{n-j-t-1})(d^{j+1}-d^p)\\
 &=& (t-p)(d^{n-t+1}-d^{n-t}) + d^{n-2t+p}-d^t.
\end{eqnarray*}

{\bf Case 3.} When $t \leq p$ (which is $\leq n-t-1$), set
%$\alpha_w=1$ for $p \leq j(w) \leq n-t-1$ and all the other
%$\alpha_w$ and $\beta_{\mv u,w}$ to be zero. 
\begin{equation}
\alpha_w =
 \begin{cases}
     1 & p \leq j(w) \leq n-t-1\\
     0 & \text{otherwise}
 \end{cases}
\end{equation}
and all the $\beta_{\mv u,w}$ to be zero.
Again, the feasibility of
the constraints (DC-1) is easy to verify. The contribution
to the objective value is 
%\[ \sum_{j=p}^{n-t-1} d^t(d^{n-j-t}-d^{n-j-t-1}) = d^{n-p}-d^t. \]
\begin{eqnarray*}
 \sum_{\substack{w\\p \leq j(w) < n-t}} d^t\alpha_w
&=& \sum_{j=p}^{n-t-1} d^t \cdot |\{w : j(w) = j\}|\\
&=& \sum_{j=p}^{n-t-1} d^t(d^{n-j-t}-d^{n-j-t-1}) \\
&=& d^{n-p}-d^t.
\end{eqnarray*}

\item
{\bf Specifying the $\gamma_{\mv u}$ and $\delta_{\mv v}$ variables.}
When $q = n-t$, set $\delta_{\mv v}=1$ for all 
$\mv v \in \bigcup_{j=n-t}^{n-1} B_j$
and all $\gamma_{\mv u}=0$. The dual-objective contribution in this case
is 
\[ \sum_{\mv v \in \bigcup_{j=n-t}^{n-1} B_j} \delta_{\mv v} = 
   |\bigcup_{j=n-t}^{n-1} B_j| = d^t - k.
\]
When $n-t+1\leq q \leq n$, define
$\delta_{\mv v}=1$ for all $\mv v \in \bigcup_{j=q}^{n-1} B_j$,
$\gamma_{\mv u}=1$ for all $\mv u$ such that 
$n-q\leq i(\mv u) \leq n-p-1$,
and all other $\delta_{\mv v}$ and $\gamma_{\mv u}$ are set to be zero.
\iffalse
\begin{equation}
\delta_{\mv v} = 
 \begin{cases}
  1 & \mv v \in \bigcup_{j=q}^{n-1} B_j\\
  0 & \text{otherwise}
 \end{cases}
\end{equation}
and
\begin{equation}
\gamma_{\mv u} = 
 \begin{cases}
  1 & n-q\leq i(\mv u) \leq n-p-1\\
  0 & \text{otherwise}
 \end{cases}
\end{equation}
\fi
From Proposition \ref{prop:q-bound}, 
the total contribution of the $\gamma_{\mv u}$ and 
$\delta_{\mv v}$ to the dual-objective is %at most
%\[
% \sum_{i=n-q}^{n-p-1}|A_i| + \min\{d^t-k, k(d^{n-q}-1)\} 
% = d^q - d^p + \min\{d^t-k, k(d^{n-q}-1)\}.
%\]
\begin{eqnarray*}
 \sum_{\substack{\mv u\\n-q\leq i(\mv u) \leq n-p-1}} \gamma_{\mv u} +
 \sum_{\mv v \in \bigcup_{j=q}^{n-1} B_j} \delta_{\mv v} 
 &=& \sum_{i=n-q}^{n-p-1}|\{\mv u : i(\mv u)=i\}| 
     + \left|\bigcup_{j=q}^{n-1} B_j\right| \\
 &\leq& \sum_{i=n-q}^{n-p-1}|A_i| + \min\{d^t-k, k(d^{n-q}-1)\} \\
 &=& d^{q} - d^p + \min\{d^t-k, k(d^{n-q}-1)\}.
\end{eqnarray*}
The feasibility of all the constraints (DC-2) is easy to verify.
\ei

Define the ``cost'' $g(k,p,q)$ to be
the total contribution of all variables to the dual-objective value.
We summarize the values of $g(k,p,q)$ in 
Figure~\ref{fig:cost-cf}.
We just proved the following.

\begin{figure*}[t]
\footnotesize
\bbox \centerline{\bf The objective value $g(k,p,q)$} \ebox
For $t \geq \left\lceil \frac n 2 \right\rceil$ and $q=n-t$,
\[
g(k,p,q) = 
   f(d^p-1)+(n-t-p)(d^{n-t+1}-d^{n-t}) - d^{n-t}+d^p + d^t-k.
\]
For $t \geq \left\lceil \frac n 2 \right\rceil$ and $q>n-t$,
\[
g(k,p,q) = 
   f(d^p-1)+(n-t-p)(d^{n-t+1}-d^{n-t}) - d^{n-t}+
   d^q + \min\{d^t-k, k(d^{n-q}-1)\}.
\]
For $p+1 \leq t \leq \left\lceil \frac n 2 \right\rceil-1$ and $q=n-t$
\[
g(k,p,q) = 
   f(d^p-1)+(t-p)(d^{n-t+1}-d^{n-t}) + d^{n+p-2t}-k.
\]
For $p+1 \leq t \leq \left\lceil \frac n 2 \right\rceil-1$ and $q>n-t$
\[
g(k,p,q) = 
   f(d^p-1)+(t-p)(d^{n-t+1}-d^{n-t}) + d^{n+p-2t}-d^t + d^q-d^p+
   \min\{d^t-k, k(d^{n-q}-1)\}.
\]
For $t \leq p$ and $q=n-t$,
\[ g(k,p,q) = f(d^p-1)+d^{n-p}-k. \]
For $t \leq p$ and $q>n-t$,
\[
g(k,p,q) = 
   f(d^p-1)+d^{n-p}-d^t+d^q-d^p+ \min\{d^t-k, k(d^{n-q}-1)\}.
\]
\caption{The dual objective value of the family of dual-feasible solutions.}
\label{fig:cost-cf}
\end{figure*}

\bthm
The above family of solutions is feasible for the dual LP \eqref{eqn:dual-LP}
with objective value equal to $g(k,p,q)$.
(Recall that, in this problem we are working on the dual constraints
for which $i(\mv u)+j(\mv v)\geq n-1$ and $i(\mv u)+j(w)\geq n-1$.)
Consequently, for the network $\log_d(N,0,m)$ to be crosstalk-free
$f$-cast wide-sense
nonblocking under the window algorithm with window size $d^t$, it is 
sufficient that
\beq
  m \geq 1+\max_{1\leq k\leq \min(f,d^t)} \min_{p,q} g(k,p,q). 
\label{eqn:sufcond2}
\eeq
\label{thm:mainlog2}
\ethm

\subsection{Selecting the best dual-feasible solution}

\begin{figure*}[t]
\footnotesize
\bbox \centerline{\bf The upper-bound $G(t,f)$} \ebox
To shorten the notations, let $r = \lfloor \log_d f \rfloor$.
\[
G(t,f) = 
 \begin{cases}
d^{n-t}[(n-t)(d-1)-1]+d^t-d^{2t-n+1}(d-1) 
& t>n/2, r \geq \max\{2t-n-2, n-t+1\}\\
f(d^{n-t-r}-1)+rd^{n-t}(d-1)-d^{n-t}+
d^{\left\lfloor \frac{r+n+1}{2} \right\rfloor}+
  f(d^{n-\left\lfloor \frac{r+n+1}{2} \right\rfloor}-1)
& t>n/2, r \leq \min\{2t-n-3, n-t\}\\
d^{n-t}[(n-t)(d-1)-1]+
d^{\left\lfloor \frac{r+n+1}{2} \right\rfloor}+
  f(d^{n-\left\lfloor \frac{r+n+1}{2} \right\rfloor}-1)
& t>n/2, n-t+1 \leq r \leq 2t-n-3\\
f(d^{n-t-r}-1)+d^{n-t}[r(d-1)-1] + d^t-d^{2t-n-2}(d-1)
& t>n/2, 2t-n-2\leq r \leq n-t\\
d^{n-t}[(n-t)(t-1)-1] + d^t
& t=n/2\\
f(d^{\left\lceil \frac{n-r-1}{2} \right\rceil}-1)
+ d^{n-\left\lceil \frac{n-r-1}{2} \right\rceil} -1
& t<n/2, r\leq n-2t$ and $f\leq d^{n-2t}(d-1)\\
f(d^{t-1}-1)+d^{n-t-1}(d^2-d+1)-1
& t<n/2, r\leq n-2t, f> d^{n-2t}(d-1)\\
f(d^{n-t-r}-1)+(2t-n-r)(d-1)d^{n-t}+ d^{2n-3t-r} -1
& t<n/2, n-2t+1\leq r \leq n-t\\
t(d-1)d^{n-t} + d^{n-2t} - 1
& t<n/2, n-t<r
 \end{cases}
\]
\caption{We show in Theorem \ref{thm:general-cf}
         that $G(t,f) \geq \max_k \min_{p,q} g(k,p,q)$}
\label{fig:G(t,f)}
\end{figure*}

The proof of the following technical lemma is similar to that
of Lemma \ref{lmm:technical1}, and thus we omit the proof.
\begin{lmm}
Let $d,n,k$ be positive integers, and $x=\lfloor \log_d k \rfloor$.
Then, the following function
\begin{equation}
 \bar h(k) = 
d^{\left\lfloor \frac{x+n+1}{2} \right\rfloor} +
k\left(d^{n-\left\lfloor \frac{x+n+1}{2} \right\rfloor}-1\right)
\label{eqn:barh(k)}
\end{equation}
is non-decreasing in $k$.
\label{lmm:technical2}
\end{lmm}

\bthm
The $\log_d(N, 0, m)$ network is crosstalk-free nonblocking under the 
window algorithm with window size $d^t$ if 
$m \geq 1+G(t,f)$ where $G(t,f)$ is defined in Figure 
\ref{fig:G(t,f)}.
\label{thm:general-cf}
\ethm
\bp
%Consider $5$ cases in the definition of $C(t,f)$.
We specify for each $k$ how to set the values
$p_k$ and $q_k$.
%And then, we verify that $c(k,p_k,q_k) \leq C(t,f)$.
The straightforward task of verifying that $g(k,p_k,q_k) \leq G(t,f)$
is mostly omitted due to space constraint.

Suppose $t>n/2$, i.e. $2t\geq n+1$.
We consider four cases as follows.
In all cases, define an integral variable $x = \lfloor \log_dk \rfloor$.

\bi
\item
{\bf Case 1.} $r \geq \max\{2t-n-2, n-t+1\}$. 

If $k \geq d^{2t-n-2}(d-1)+1$, then pick $p_k=0$ and $q_k=n-t$. 
\iffalse
Then,
\begin{eqnarray*}
&&g(k,p_k,q_k)\\
&=&(n-t)(d^{n-t+1}-d^{n-t})-d^{n-t}+1+d^t-k\\
&\leq&(n-t)(d^{n-t+1}-d^{n-t})-d^{n-t}+1+\\
&&d^t-d^{2t-n+1}(d-1)-1\\
&=&d^{n-t}[(n-t)(d-1)-1]+d^t-d^{2t-n+1}(d-1).
\end{eqnarray*}
\fi
On the other hand, if $k\leq d^{2t-n-2}(d-1)$ then we pick
$p_k=0$ and $q_k=\left\lfloor \frac{x+n+1}{2} \right\rfloor > n-t$.
Note that $kd^{n-q_k} \leq d^t$.
Thus, recall from Lemma \ref{lmm:technical2} that the function
$\bar h(k)$ defined in \eqref{eqn:barh(k)} is non-decreasing in $k$, we
have
\begin{eqnarray*}
g(k,p_k,q_k)
&=&(n-t)(d^{n-t+1}-d^{n-t})-d^{n-t}+
d^{\left\lfloor \frac{x+n+1}{2} \right\rfloor} +
\min\left\{d^t-k, k\left(d^{n-\left\lfloor \frac{x+n+1}{2} \right\rfloor}-
1\right)\right\}\\
&=&(n-t)(d^{n-t+1}-d^{n-t})-d^{n-t}+
d^{\left\lfloor \frac{x+n+1}{2} \right\rfloor} +
k\left(d^{n-\left\lfloor \frac{x+n+1}{2} \right\rfloor}\right)\\
&=&(n-t)(d^{n-t+1}-d^{n-t})-d^{n-t} + \bar h(k)\\
&\leq&(n-t)(d^{n-t+1}-d^{n-t})-d^{n-t} + \bar h(d^{2t-n-2}(d-1))\\
&=&d^{n-t}[(n-t)(d-1)-1]+d^t-d^{2t-n+1}(d-1).
\end{eqnarray*}
%The function
%\[ \bar h(k) = d^{\left\lfloor \frac{x+n+1}{2} \right\rfloor} +
%k\left(d^{n-\left\lfloor \frac{x+n+1}{2} \right\rfloor}-1\right)
%\]
%is nondecreasing in integral $k$, which is at most $d^{2t-n-2}(d-1)$.
%And, when $k=d^{2t-n-2}(d-1)$ we have $x=2t-n-2$ in which case
%$\left\lfloor \frac{x+n+1}{2} \right\rfloor = t-1$.
%Hence,
%\begin{eqnarray*}
%g(k,p_k,q_k)
%&\leq& (n-t)(d^{n-t+1}-d^{n-t})-d^{n-t}+
%d^{t-1} + d^{2t-n-2}(d-1)\left(d^{n-t+1}-1\right)\\
%&=&d^{n-t}[(n-t)(d-1)-1]+d^t-d^{2t-n+1}(d-1).
%\end{eqnarray*}
%In summary, in case 1 when $r \geq \max\{2t-n-2, n-t+1\}$ we always
%have 
%\[ g(k,p_k,q_k) \leq d^{n-t}[(n-t)(d-1)-1]+d^t-d^{2t-n+1}(d-1). \]

\item
{\bf Case 2.} $r \leq \min\{2t-n-3, n-t\}$.
set $p_k=n-t-r \leq t-1$ and 
$q_k=\left\lfloor \frac{x+n+1}{2} \right\rfloor > n-t$.
\iffalse
In this case, we set $p_k=n-t-r \leq t-1$ and 
$q_k=\left\lfloor \frac{x+n+1}{2} \right\rfloor > n-t$.
Similar to the above analysis, we obtain
\begin{eqnarray*}
&& g(k, p_k, q_k) \\
&=& f(d^{n-t-r}-1)+rd^{n-t}(d-1)-d^{n-t}+\\
&&
d^{\left\lfloor \frac{x+n+1}{2} \right\rfloor}+
  k(d^{n-\left\lfloor \frac{x+n+1}{2} \right\rfloor}-1)\\
&=&f(d^{n-t-r}-1)+rd^{n-t}(d-1)-d^{n-t}+\bar h(k)\\
&\leq&f(d^{n-t-r}-1)+rd^{n-t}(d-1)-d^{n-t}+\bar h(f)\\
&=& f(d^{n-t-r}-1)+rd^{n-t}(d-1)-d^{n-t}+\\
&&
d^{\left\lfloor \frac{r+n+1}{2} \right\rfloor}+
  f(d^{n-\left\lfloor \frac{r+n+1}{2} \right\rfloor}-1).
\end{eqnarray*}
For the inequality, we used the fact that the function $\bar h(k)$ is
non-decreasing in $k$, and $k\leq f$.
\fi
\item
{\bf Case 3.} $n-t+1 \leq r \leq 2t-n-3$.
This is case we  set $p_k=0$ and 
$q_k=\left\lfloor \frac{x+n+1}{2} \right\rfloor > n-t$.
%Then,
%\begin{multline*}
% g(k,p_k,q_k) \leq \\d^{n-t}[(n-t)(d-1)-1]+
%d^{\left\lfloor \frac{r+n+1}{2} \right\rfloor}+
%  f(d^{n-\left\lfloor \frac{r+n+1}{2} \right\rfloor}-1).
%\end{multline*}

\item
{\bf Case 4.} $2t-n-2\leq r \leq n-t$. In this case we set
$p_k=n-t-r$ and consider two sub-cases as in case 1:
$k\geq d^{2t-n-2}(d-1)+1$ or $k\leq d^{2t-n-2}(d-1)$.
%The eventual bound is
%\begin{multline*}
%g(k,p_k,q_k) \leq \\f(d^{n-t-r}-1)+d^{n-t}[r(d-1)-1] +
%d^t-d^{2t-n-2}(d-1).
%\end{multline*}
\ei

When $t=n/2$, we simply set $q_k=n-t$ and $p_k=0$.
Then, 
\[ g(k,p_k,q_k) = d^{n-t}[(n-t)(d-1)-1]+d^t. \]

Finally, suppose $t<n/2$.
We will always pick $p_k=q-t$ in this situation.
Also consider four cases:

\bi
\item
{\bf Case 1.} $r\leq n-2t$ and $f\leq d^{n-2t}(d-1)$,
set $p_k=\left\lceil \frac{n-r-1}{2} \right\rceil \geq t$.
%In this case,
%\[ g(k, p_k,q_k) = f(d^{\left\lceil \frac{n-r-1}{2} \right\rceil}-1)
%                   + d^{n-\left\lceil \frac{n-r-1}{2} \right\rceil} -1.
%\]

\item
{\bf Case 2.} $r\leq n-2t$ and $f> d^{n-2t}(d-1)$,
set $p_k=t-1$.  
%In this case,
%\[ g(k, p_k,q_k) = f(d^{t-1}-1)+d^{n-t-1}(d^2-d+1)-1.  \]

\item
{\bf Case 3.} $n-2t+1\leq r \leq n-t$,
set $p_k=n-t-r$.  
%In this case,
%\begin{multline*}
% g(k, p_k,q_k) = \\
% f(d^{n-t-r}-1)+(2t-n-r)(d-1)d^{n-t}+ d^{2n-3t-r} -1.  
%\end{multline*}

\item
{\bf Case 4.} $n-t<r$,
set $p_k=0$.  
%In this case,
%\[ g(k, p_k,q_k) = t(d-1)d^{n-t} + d^{n-2t} - 1. \]
\ei
\ep

\bcor[Theorems $1$ in \cite{ngo-nguyen-ha:gc08}]
The $d$-ary multi-log network $\log_d(N,0,m)$
is crosstalk-free wide-sense nonblocking with respect to the window algorithm
with window size $d^t$ if 
\[ 
m \geq 
\begin{cases}
  d^{n-2t}+td^{n-t}(d-1) & t<n/2\\
 d^{n-t}[(n-t)(d-1)-1]+d^t+1 & t=n/2\\
 d^{n-t}[(n-t)(d-1)-1]+&\\
 d^t-d^{2t-n-2}(d-1)+1 & t>n/2.
\end{cases}
\]
\label{cor:ngo-nguyen-ha:gc08}
\ecor

%\input{hardness}
%\input{conclusions}

%\section*{Acknowledgements}
%
%Hung Q. Ngo was supported in part by NSF CAREER Award CCF-0347565.
%Atri Rudra was supported in part by NSF CAREER Award CCF-0844796.

%%%%%%%%%%%%%%%%%%%%%%%%%%%%%%%%%%%%%%%%%%%%%%%%%%%%%%%%%%%%%%%%%%%%%%%%%%%%%%%
%% The Bib

%%      BIBLIOGRAPHY
%\iffalse
{
  %% Options include :
  %%  abbrv, alpha, ieeetr, siam, acm, apalike, plain, unsrt
  % \fontsize{9}{11pt}\selectfont
  %\bibliographystyle{acm}
  %\bibliography{main}
  %\def\DIR#1{\gdef\@DIR{#1}}

  % \DIR{/u0/csefaculty/hungngo/Papers/BIB}
  %\DIR{/Users/hungngo/Dropbox/BIB}
  % \DIR{.}

%\bibliography{\@DIR/generalMath,\@DIR/HungNgo,\@DIR/WDM_OXC,\@DIR/switchingNet,\@DIR/LogNetRNB,\@DIR/approximation}

\def\polhk#1{\setbox0=\hbox{#1}{\ooalign{\hidewidth
  \lower1.5ex\hbox{`}\hidewidth\crcr\unhbox0}}} \def\cprime{$'$}

}
%\fi

\end{document}

%% file: clos.pdftex_t
\begin{picture}(0,0)%
\epsfig{file=clos.pdf}%
\end{picture}%
\setlength{\unitlength}{1618sp}%
\begingroup\makeatletter\ifx\SetFigFont\undefined%
\gdef\SetFigFont#1#2#3#4#5{%
  \reset@font\fontsize{#1}{#2pt}%
  \fontfamily{#3}\fontseries{#4}\fontshape{#5}%
  \selectfont}%
\fi\endgroup%
\begin{picture}(8863,4224)(751,-4573)
\put(9076,-3736){\makebox(0,0)[lb]{\smash{{\SetFigFont{6}{7.2}{\rmdefault}{\mddefault}{\updefault}$n_2$}}}}
\put(1951,-1336){\makebox(0,0)[lb]{\smash{{\SetFigFont{6}{7.2}{\rmdefault}{\mddefault}{\updefault}$I_1$}}}}
\put(4951,-736){\makebox(0,0)[lb]{\smash{{\SetFigFont{6}{7.2}{\rmdefault}{\mddefault}{\updefault}$M_1$}}}}
\put(4951,-4336){\makebox(0,0)[lb]{\smash{{\SetFigFont{6}{7.2}{\rmdefault}{\mddefault}{\updefault}$M_m$}}}}
\put(7951,-1336){\makebox(0,0)[lb]{\smash{{\SetFigFont{6}{7.2}{\rmdefault}{\mddefault}{\updefault}$O_1$}}}}
\put(1876,-3736){\makebox(0,0)[lb]{\smash{{\SetFigFont{6}{7.2}{\rmdefault}{\mddefault}{\updefault}$I_{r_1}$}}}}
\put(7876,-3736){\makebox(0,0)[lb]{\smash{{\SetFigFont{6}{7.2}{\rmdefault}{\mddefault}{\updefault}$O_{r_2}$}}}}
\put(751,-3736){\makebox(0,0)[lb]{\smash{{\SetFigFont{6}{7.2}{\rmdefault}{\mddefault}{\updefault}$n_1$}}}}
\put(751,-1336){\makebox(0,0)[lb]{\smash{{\SetFigFont{6}{7.2}{\rmdefault}{\mddefault}{\updefault}$n_1$}}}}
\put(2626,-1336){\makebox(0,0)[lb]{\smash{{\SetFigFont{6}{7.2}{\rmdefault}{\mddefault}{\updefault}$m$}}}}
\put(2626,-3586){\makebox(0,0)[lb]{\smash{{\SetFigFont{6}{7.2}{\rmdefault}{\mddefault}{\updefault}$m$}}}}
\put(4051,-961){\makebox(0,0)[lb]{\smash{{\SetFigFont{6}{7.2}{\rmdefault}{\mddefault}{\updefault}$r_1$}}}}
\put(4051,-4036){\makebox(0,0)[lb]{\smash{{\SetFigFont{6}{7.2}{\rmdefault}{\mddefault}{\updefault}$r_1$}}}}
\put(5776,-961){\makebox(0,0)[lb]{\smash{{\SetFigFont{6}{7.2}{\rmdefault}{\mddefault}{\updefault}$r_2$}}}}
\put(5776,-4036){\makebox(0,0)[lb]{\smash{{\SetFigFont{6}{7.2}{\rmdefault}{\mddefault}{\updefault}$r_2$}}}}
\put(9076,-1336){\makebox(0,0)[lb]{\smash{{\SetFigFont{6}{7.2}{\rmdefault}{\mddefault}{\updefault}$n_2$}}}}
\end{picture}%